\documentclass[twoside,twocolumn,9pt]{article}
\usepackage{extsizes}
\usepackage[super,sort&compress,comma]{natbib} 
\usepackage[version=3]{mhchem}
\usepackage[left=1.5cm, right=1.5cm, top=1.785cm, bottom=2.0cm]{geometry}
\usepackage{balance}
\usepackage{mathptmx}
\usepackage{sectsty}
\usepackage{graphicx} 
\usepackage{stackengine}
\usepackage{lastpage}
\usepackage[format=plain,justification=justified,singlelinecheck=false,font={stretch=1.125,small,sf},labelfont=bf,labelsep=space]{caption}
\usepackage{float}
\usepackage{fancyhdr}
\usepackage{fnpos}
\usepackage[english]{babel}
\addto{\captionsenglish}{%
  \renewcommand{\refname}{Notes and references}
}
\usepackage{amssymb}
\usepackage{array}
\usepackage{droidsans}
\usepackage{charter}
\usepackage[T1]{fontenc}
\usepackage[usenames,dvipsnames]{xcolor}
\usepackage{setspace}
\usepackage[compact]{titlesec}
\usepackage{hyperref}
\usepackage{epstopdf}%This line makes .eps figures into .pdf - please comment out if not required.

\definecolor{cream}{RGB}{222,217,201}

\usepackage{bm}
\begin{document}

\pagestyle{fancy}
\thispagestyle{plain}
\fancypagestyle{plain}{
%%%HEADER%%%
\renewcommand{\headrulewidth}{0pt}
}
%%%END OF HEADER%%%

%%%PAGE SETUP - Please do not change any commands within this section%%%
\makeFNbottom
\makeatletter
\renewcommand\LARGE{\@setfontsize\LARGE{15pt}{17}}
\renewcommand\Large{\@setfontsize\Large{12pt}{14}}
\renewcommand\large{\@setfontsize\large{10pt}{12}}
\renewcommand\footnotesize{\@setfontsize\footnotesize{7pt}{10}}
\makeatother

\renewcommand{\thefootnote}{\fnsymbol{footnote}}
\renewcommand\footnoterule{\vspace*{1pt}% 
\color{cream}\hrule width 3.5in height 0.4pt \color{black}\vspace*{5pt}} 
\setcounter{secnumdepth}{5}

\makeatletter 
\renewcommand\@biblabel[1]{#1}            
\renewcommand\@makefntext[1]% 
{\noindent\makebox[0pt][r]{\@thefnmark\,}#1}
\makeatother 
\renewcommand{\figurename}{\small{Fig.}~}
\sectionfont{\sffamily\Large}
\subsectionfont{\normalsize}
\subsubsectionfont{\bf}
\setstretch{1.125} %In particular, please do not alter this line.
\setlength{\skip\footins}{0.8cm}
\setlength{\footnotesep}{0.25cm}
\setlength{\jot}{10pt}
\titlespacing*{\section}{0pt}{4pt}{4pt}
\titlespacing*{\subsection}{0pt}{15pt}{1pt}
%%%END OF PAGE SETUP%%%

%%%FOOTER%%%
\fancyfoot{}
\fancyfoot[LO,RE]{\vspace{-7.1pt}\includegraphics[height=9pt]{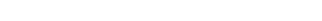}}
\fancyfoot[CO]{\vspace{-7.1pt}\hspace{13.2cm}\includegraphics{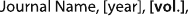}}
\fancyfoot[CE]{\vspace{-7.2pt}\hspace{-14.2cm}\includegraphics{head_foot/RF}}
\fancyfoot[RO]{\footnotesize{\sffamily{1--\pageref{LastPage} ~\textbar  \hspace{2pt}\thepage}}}
\fancyfoot[LE]{\footnotesize{\sffamily{\thepage~\textbar\hspace{3.45cm} 1--\pageref{LastPage}}}}
\fancyhead{}
\renewcommand{\headrulewidth}{0pt} 
\renewcommand{\footrulewidth}{0pt}
\setlength{\arrayrulewidth}{1pt}
\setlength{\columnsep}{6.5mm}
\setlength\bibsep{1pt}
%%%END OF FOOTER%%%

%%%FIGURE SETUP - please do not change any commands within this section%%%
\makeatletter 
\newlength{\figrulesep} 
\setlength{\figrulesep}{0.5\textfloatsep} 

\newcommand{\topfigrule}{\vspace*{-1pt}% 
\noindent{\color{cream}\rule[-\figrulesep]{\columnwidth}{1.5pt}} }

\newcommand{\botfigrule}{\vspace*{-2pt}% 
\noindent{\color{cream}\rule[\figrulesep]{\columnwidth}{1.5pt}} }

\newcommand{\dblfigrule}{\vspace*{-1pt}% 
\noindent{\color{cream}\rule[-\figrulesep]{\textwidth}{1.5pt}} }

\makeatother
%%%END OF FIGURE SETUP%%%

%%%TITLE, AUTHORS AND ABSTRACT%%%
\twocolumn[
  \begin{@twocolumnfalse}
{\includegraphics[height=30pt]{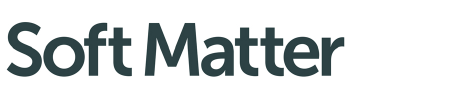}\hfill\raisebox{0pt}[0pt][0pt]{\includegraphics[height=55pt]{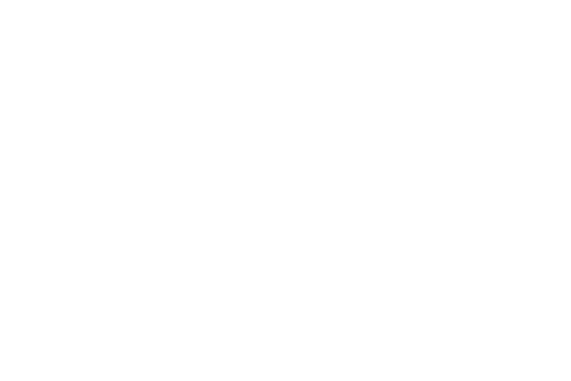}}\\[1ex]
\includegraphics[width=18.5cm]{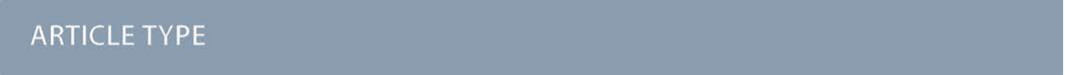}}\par
\vspace{1em}
\sffamily
\begin{tabular}{m{4.5cm} p{13.5cm} }

\includegraphics{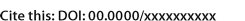} & \noindent\LARGE{\textbf{ 
Amoeboid propulsion of active solid bodies, vesicles and droplets: a comparison
}} \\%Article title goes here instead of the text "This is the title"
\vspace{0.3cm} & \vspace{0.3cm} \\

 & \noindent\large{Reiner Kree,$^{\ast}$\textit{$^{a}$} Annette Zippelius,$^{\ast}$\textit{$^{a}$}} \\%Author names go here instead of "Full name", etc.

\includegraphics{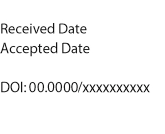} & \noindent\normalsize{
We present a unified discussion of three types of near-spherical  amoeboid microswimmers, driven by periodic, axially symmetric, achiral deformations (swim strokes): a solid deformable body, a vesicle with incompressible fluid membrane, and a droplet. Minimal models are used, which characterize the swimmer type only by boundary conditions. We calculate the swimming velocities, the dissipated power and the Lighthill efficiencies within a second order perturbation expansion in the small deformation amplitudes. %Our approach uses spherical harmonics to represent surface deformations and a system of general solutions of the Stokes equation based on vector spherical harmonics. 
For solid bodies, we reproduce older results by Lighthill and Blake, for vesicles and for droplets we add new results. The unified approach allows for a detailed comparison between the three types of microswimmers. We present such comparisons for swim strokes made up of spherical harmonics of adjacent orders $l$ and $l+1$, as well as for a manifold of swim strokes, made up of spherical harmonics up to order $l=4$, which respect volume- and surface-incompressibility. This manifold is two-dimensional, which allows to present swimming velocities and efficiencies in compact graphical form.  
In a race in which each swimmer can choose the stroke that maximizes its speed, the droplet always comes in first, the vesicle comes in second, while the particle finishes third. However, if the three swimmers perform the same stroke, other order of rankings become possible. 
The maximum of the total efficiency of a droplet is greater than that of a vesicle if the internal dissipation is small. The efficiency of the solid body turns out to be typically two orders of magnitude smaller than that of vesicles and droplets. Optimizing the Lighthill efficiency and optimizing the swimming velocity result in different optimal swim strokes
} \\%The abstrast goes here instead of the text "The abstract should be..."

\end{tabular}

 \end{@twocolumnfalse} \vspace{0.6cm}
]
%%%END OF TITLE, AUTHORS AND ABSTRACT%%%

%%%FONT SETUP - please do not change any commands within this section
\renewcommand*\rmdefault{bch}\normalfont\upshape
\rmfamily
\section*{}
\vspace{-1cm}

%%%FOOTNOTES%%%

\footnotetext{\textit{$^{a}$~Institute of Theoretical Physics, Georg-August University, Friedrich-Hund Pl. 1,  D-37077 Göttingen, Germany. E-mail: rkree1@phys.uni-goettingen.de}}
%\footnotetext{\textit{$^{b}$~Address, Address, Town, Country. }}

%Please use \dag to cite the ESI in the main text of the article.
%If you article does not have ESI please remove the the \dag symbol from the title and the footnotetext below.
%\footnotetext{\dag~Supplementary Information available: [details of any supplementary information available should be included here]. See DOI: 10.1039/cXsm00000x/}
%additional addresses can be cited as above using the lower-case letters, c, d, e... If all authors are from the same address, no letter is required

%\footnotetext{\ddag~Additional footnotes to the title and authors can be included \textit{e.g.}\ `Present address:' or `These authors contributed equally to this work' as above using the symbols: \ddag, \textsection, and \P. Please place the appropriate symbol next to the author's name and include a \texttt{\textbackslash footnotetext} entry in the the correct place in the list.}

%%%END OF FOOTNOTES%%%

%%%MAIN TEXT%%%%

\section{Introduction}

 Self-propelling microorganisms have developed a wide variety of mechanisms to exert driving forces on different environments~\cite{Paluch2016,Othmer2019}. The crawling of cells on a solid substrate, for example, is accomplished by specific adhesion sites. This mechanism has been extensively studied since the work of Abercrombie ~\cite{Abercrombie}.  
It has also been shown that self-propulsion in
confinements such as solid micro-channels is possible without adhesion sites ~\cite{Bergert,Paluch2016}.
If microorganisms move in fluid environments, the transmitted
forces are dominated by {\it viscous drag}, as the Reynolds number of the generated flow is very small. Biological cells create propelling tractions by time-asymmetric sequences of morphological changes, known as swim strokes, either through the beating of short (\textit{cilia}) or longer (\textit{flagella}) active filaments, or by shape deformations of the entire cell (amoeboid motion).  On hydrodynamic scales, the beating of a carpet of cilia covering the surface of the organism is equivalent to small shape changes, as noticed by Lighthill ~\cite{Lighthill} and Blake ~\cite{Blake}. A number of eukaryotic cells were previously thought to need an underlying substrate to move (crawl) using shape deformations, but growing evidence now suggests that many types of cells are capable of migrating without focal adhesion, enabling them to swim just as effectively as they crawl.
Among the organisms which have been found to move by shape changes,
are not just amoebae ~\cite{Ebata2024}, but also neutrophils~\cite{Barry}, mutants of Dicties~\cite{Barry,Bodenschatz} and Euglenids~\cite{Arroyo}. In fact, some of these organisms can change between different modes of swimming strategies
depending on the environment. Neutrophils and Dicties were shown to
perform chemotaxis in solution and, under appropriate environmental
conditions, crawl on solid surfaces as well~\cite{Barry,Bodenschatz}.

Inspired by the swimming of microorganisms, great efforts have been made to construct self-propelled or field-propelled synthetic microrobots, which can serve many different purposes in applications.  
These attempts start from a solid body, a vesicle, or a droplet, driven by a variety of active mechanisms \cite{Yi2009, Boukellal2004, Shah2014, Spatz2017, Palagi2018, Maroto2019, Gompper2019, Vukturi2020}. 
 
In the present work, we focus on amoeboid motion. Theoretical approaches~\cite{Jones} help to understand the fundamental hydromechanical principles of these types of drive.
The problem can be decomposed into two parts. First, a variety of active drives\cite{Gompper2019, Vukturi2020, Kree2021, Kree2022, Zippelius2021, Sprenger2020} (cytosceletons, internalized active microswimmers , active fibers e.t.c.) are studied, which lead to deformations of the surface, and second, the propulsion and the fluid flows due to these deformations have to be calculated. Here, we concentrate on the second part of the problem, which builds on the work of Taylor~\cite{Taylor}, who computed the swimming speed of an inextensible sheet performing wavelike oscillations. Lighthill~\cite{Lighthill} extended Taylor`s work to a sphere with small surface deformations (squirmer)
and computed its swimming velocity and energetic efficiency. Later, Lighthill`s work was corrected and extended by Blake~\cite{Blake}, and it has been analyzed in great detail in the following years
~\cite{Lauga2009, Pak2014, Shapere1,Felderhof1994_I,Felderhof1994_II}. Energy
dissipation and efficiency have been calculated and optimized with
respect to possible shape deformations ~\cite{Shapere2,Felderhof2014}.
A two-dimensional variant of the model has been applied to Eukaryotic
cells ~\cite{Wang2018}, which were experimentally observed to swim with only viscous traction.  

As a bridge between the first and second part of the problem, models that start from active tractions localized at the surface  \cite{Kree2023, Misbah2013} have
also been considered. Deformations of droplets driven by Marangoni flow have been analyzed in several studies\cite{Morozov2018, Yoshinaga2014}. 
Farutin et al.~\cite{Misbah2013} model an
amoeboid swimmer as a vesicle driven by active membrane forces that cause
shape deformations and self-propulsion. 
More strongly coarse-grained models~\cite{Ohta2017}
discuss the motion of a deformable particle in terms of coupled
equations for the flow and a quadrupolar order parameter, whereas more
detailed computational models~\cite{Campbell,Wang2016} aim to mimic specific biological processes such as
pseudopod formation, including biochemical reactions in the membrane.

In the following sections, we present a unified approach to minimal models of single, nearly spherical amoeboid microswimmers of three different types: (i) a deformable solid, (ii) a vesicle with a fluid, incompressible membrane, and (iii) a droplet. The swimmers move in a Newtonian fluid, and the vesicle and droplet are filled with another Newtonian fluid. For all three types, we calculate the swimming velocity and the efficiency due to \textit{given} swim strokes using the same analytical perturbation method applicable for small swim stroke amplitudes\cite{Taylor1934}\cite{Seifert1999}\cite{Rallison1980}. Although we do not treat it in detail, it will be obvious that the method also gives the fluid flow fields. It reproduces the results of Lighthill~\cite{Lighthill} and Blake~\cite{Blake} for a squirmer, and it complements the results of Farutin et al. ~\cite{Misbah2013}, who consider given membrane tractions, which produce the swim strokes. The results for active droplets are new to the best of our knowledge. Subsequently, we compare the swimming velocities and efficiencies for given swim strokes between all three types of amoeboid swimmers. The deformations of the sphere during a swim stroke are constrained by volume incompressibility. To include the vesicle in the comparisons, they also have to obey the surface incompressibility constraint. 

The models of amoeboid swimmers are introduced in the next section. The perturbative solution for the swimming velocities up to second order is presented in Sec. \ref{sec:perturbationExpansion}, and the dissipation due to viscous fluid flow is calculated in Sec. \ref{sec:efficiency}. In Section \ref{sec:comparisons}, we compare the velocities and the efficiencies of the three types of particles for periodic swimming strokes. In the simplest case, the average swimming velocity does not depend on the time course of the deformations. Beyond this case, we construct a two-parameter manifold of simple harmonic deformations. Within this manifold, we compare the velocities and efficiencies for all possible strokes and identify the corresponding optimal swim strokes. We also discuss the dependence of these quantities on the viscosities of the ambient and internal fluids.  Our main results are summarized in Section \ref{sec:summary}. Some more technical points of the calculations are delegated to the Appendices \ref{append:droplet_first_order}--\ref{app:swim-strokes}.

\section{Models}
\label{model}

We study the self-propulsion resulting from time-varying shape deformations of a microorganism or an artificial microswimmer, which can be modeled in one of three ways: as a deformable solid body, as a vesicle, or as a fluid droplet.  These systems, which are all referred to as "particles" in the following, are submerged in a Newtonian fluid and are neutrally buoyant. The unit of mass is chosen such that the mass density $\rho$ of the materials is  1. Each particle occupies a volume $V$ with a smooth boundary $\partial V$. In the absence of particles, the ambient fluid is at rest in the laboratory frame. For low Reynolds number, the outer flow field obeys the Stokes equation
\begin{equation}
\label{eq:stokes}
\nabla\cdot\bm{\sigma}=\eta\nabla^2\bm{v}-\nabla p=0 , 
\end{equation}
supplemented by the incompressibility condition $\nabla\cdot\bm{v}=0$.  The viscosity of the external fluid is denoted by $\eta$, 
and its stress tensor $\bm{\sigma}$ is given by the
Cartesian components
$\sigma_{ij}=-p\delta_{ij}+\eta(\partial_iv_j+\partial_jv_i) = -p\delta_{ij} + \sigma^{visc}_{ij}$, with
the pressure $p$ determined from incompressibility.
\begin{figure}
    \centering
    \includegraphics[width=0.9\linewidth]{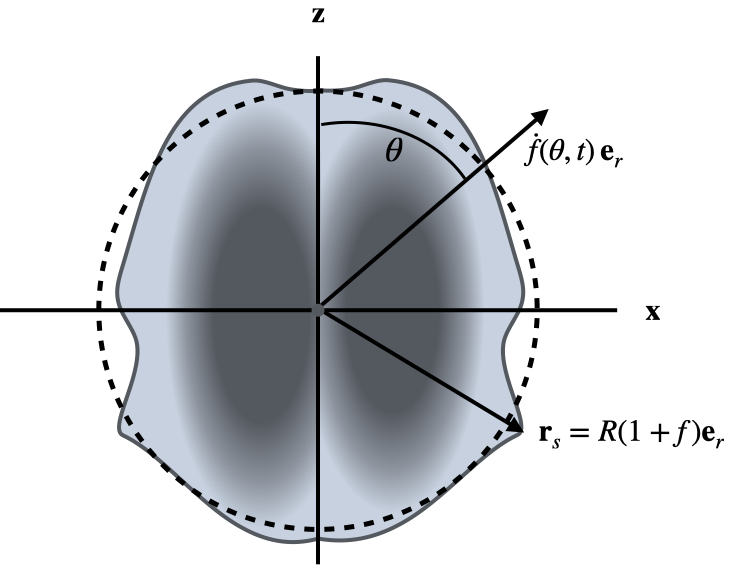}
    \caption{Illustration of setup and notation. The (uniaxial and achiral) shape changes ($d\bm{r}_s/dt=\dot{f}(\theta, t)\bm{e}_r$) are assumed to be radial with respect to the center of a reference sphere of radius $R$ (dashed line), which has the same volume as the particle. The interior can be a solid or a Newtonian fluid. For a droplet, the interface is a geometric surface, whereas for a vesicle it represents an incompressible fluid membrane.} 
    \label{fig:amoeboid-scetch}
\end{figure}

The particles are driven by small, time-dependent deformations of nearly
spherical shapes. Such deformations, caused by active forces,
have been calculated for vesicles~\cite{Misbah2013} and
droplets~\cite{Kree2023}. Here we take a different point of view: we consider the
time-dependent deformations as given and compute the corresponding propulsion of the particle. The deviations of the shape of a particle from a reference sphere, which has the same volume as the particle,  are described by a function $f$ (see Fig. \ref{fig:amoeboid-scetch}), which determines the positions $\bm{r}_s$ of surface points via the relationship

\begin{equation}
    \bm{r}_s=R(1+f)\bm{e}_r. 
    \label{eq:paramdeformation}
\end{equation}

 We use $R$ as the unit of length. The driving, which we will study in this work, is given by shape changes $\dot{f}\bm{e}_r$ (the dot denotes the time derivative), which are restricted to radial directions with respect to the center of the reference sphere. We therefore call this point the center of deformation in the following.  The radial drive can be studied analytically for all three types of particles by using a perturbation expansion, as will be shown below. For simplicity, we also restrict our considerations to uniaxial (and achiral) deformations $f(\theta, t)$. This assumption is easily lifted, but the uniaxial case is
sufficient to quantitatively compare the properties of amoeboid translational motion of the different particles. Our aim is to calculate the propulsion velocities $\bm{U}$ and the dissipated powers $\dot{E}$, which result from given deformations $f(t)$.   
We can expand the function $f(\theta, t)$ in Legendre polynomials

\begin{equation}
  f(t, \theta)=\sum_{l\neq 1}f_l(t)P_l(\cos\theta),
\end{equation}
and characterize the deformations by the amplitudes $f_l$. 
The sum does not include the term $l=1$ that would correspond to rigid body motion. 
For all three types of particles, the $f_l$ are restricted by the condition that the particle volume remains constant,  as the interior fluid or solid is taken to be incompressible. 
The time dependence of the amplitudes $f_l(t)$ has to be restricted to slow dynamics. More precisely we require the product of the Reynolds number and the Strouhal number to be small so that the partial time derivative in the Navier-Stokes equation can be neglected\cite{Kim2005}.  

Solid bodies, vesicles, and droplets are
modeled in the simplest possible way: We ignore the rheology of the
interface and instead formulate appropriate boundary conditions which
are presented below. The interior of vesicles and droplets consists of a Newtonian fluid with the same density as the ambient fluid, but in general with a different viscosity $\lambda\eta$. The inner flow will contribute to the energy that is dissipated during the motion. For the solid, we do not include inner dissipation mechanisms because this would go beyond the scope of a hydrodynamical model and leaves many possibilities. As a consequence, the dissipated energy calculated from the outer flow is only a lower bound to the total dissipation of the solid particle.  

%In the following, we concentrate on the second order of the expansion of $\bm{U}$ and $\dot{E}$. If necessary, the results can be %straightforwardly extended to the complete inner and outer flow fields in this order.  

All calculations will be performed in a  frame , in which the center
of deformation is at rest (CODF), because it is in this frame that the shapes $f(\theta,t)$ are defined. In the CODF, the flow field
$\bm{v}(\bm{r},t)$ obeys the boundary condition
\begin{equation}
\label{eq:BCinfinity}
    \lim_{r\to\infty}\bm{v}(\bm{r},t)=-\bm{U} 
\end{equation}
at infinity. The swimming velocity $\bm{U}$ is defined here as the velocity of the center of deformation and it points  in the z-direction, due to the uniaxial symmetry, i.e. $\bm{U}=U\bm{e}_z$. We could have chosen any other point of a deformable particle to define its velocity. From the perspective of dynamics, the center of mass is particularly emphasized.  In the CODF, it moves with a non-vanishing velocity $\bm{v}_{cm}$  that is
easily calculated from the deformations $f(t)$,
\begin{equation}
\label{eq:vcm}
  \bm{v}_{cm}=\frac{3}{4\pi}\frac{d}{dt}\int dV \bm{r}=\frac{3}{2}\bm{e}_z
  \int_{-1}^1 dz\, z (1+f)^3 \dot{f},
\end{equation}
and does not depend on the type of particle. The velocity of the center of mass $\bm{U}_{cm}$ in the laboratory frame is then given by $\bm{U}_{cm}=\bm{U}+\bm{v}_{cm}$. 

The models for the particles are now completed by specifying the boundary conditions, which uniquely determine the flow field.

\subsection{Deformable solid body}
For a solid particle, we use no-slip boundary conditions on its surface $\partial V$, implying
\begin{equation}\label{eq:BCsolid}
\bm{v}(\bm{r},t)=\frac{d\bm{r}_s}{dt} \quad \text{for} \quad \bm{r}\in \partial V.
\end{equation}
Together with Eq.(\ref{eq:BCinfinity}), this determines the outer flow uniquely.

\subsection{Vesicle}
In the simplest model of a fluid membrane, we  consider only the
local inextensibility, which means that the surface divergence of the flow vanishes, that is $\nabla_s\cdot \bm{v}(\bm{r}_s)=0$ for
every point $\bm{r}_s$ on $\partial V$. We can rewrite this constraint in terms of viscous tractions
$t_j^{vis}= n_i\sigma^{visc}_{ij}$, 
\begin{equation}
  \label{eq:BCV1}
  \nabla_s\cdot \bm{v}(\bm{r})=\bm{t}^{vis}(\bm{r})\cdot \bm{n}=0,
  \quad \bm{r}\in \partial V.
 \end{equation} 
 where $\bm{n}$ is the outward normal to the surface.

 The shape evolution of the vesicle is described by a level set function 
$H(r, \theta,t)=r-r_s(\theta, t)=0$ for all $t$, which implies the kinematic equation
\begin{equation}
  \label{eq:BCV2}
\frac{dH}{dt}=\frac{\partial H}{\partial t}+\bm{v}\cdot\nabla H=
-\frac{\partial f}{\partial t}+\bm{v}\cdot\nabla H=0.
\end{equation}

The boundary condition (\ref{eq:BCV1}) together with the kinematic
equation (\ref{eq:BCV2}) and the far-field asymptotics Eq. (\ref{eq:BCinfinity}) uniquely determine the outer flow field $\bm{v}$ and the swimming velocity $\bm{U}$ for given $f_l$ and $\dot{f}_l$. 
To calculate the dissipated energy, one also needs the flow inside the vesicle, which will be denoted by $\bm{V}$. It obeys the same boundary conditions on the membrane as the flow of the surrounding fluid.

In contrast to particles and droplets, the deformations $f_l$ of vesicles are further restricted because their surface area
\begin{equation}
    %A =  2\pi \int_{-\pi}^\pi\left| \frac{\partial \bm{r}_s}{\partial\theta}\times\frac{\partial \bm{r}_s}{\partial\phi} \right| d\theta
    A = 2\pi\int_{-1}^{1}(1+f) \sqrt{(1+f)^2+ f'^2} dz, 
\end{equation}
with $ z=\cos\theta$,
remains constant. 
Instead of $A$, we will use the excess area $\Delta$, defined by the relation $A=4\pi R^2(1+\Delta)$,  to express this restriction.  \cite{Seifert1999}.

\subsection{Droplet}
The droplet consists of an incompressible Newtonian fluid
with viscosity $\lambda\eta$.  The internal and external
fluids are completely immiscible and are therefore separated by a
sharp interface characterized by a constant surface tension.
%\gamma brauchen wir nicht
We use lower case symbols for the exterior and upper case symbols for the interior of the droplet. The inner flow field (which is again denoted by $\bm{V}$) has to remain finite at the center of deformation . 

The boundary conditions require continuity of the flow across the interface
\begin{equation}\label{eq:BC3}
  \bm{v}(\bm{r})=\bm{V}(\bm{r}),\qquad \bm{r}\in \partial V
\end{equation}
as well as continuity of the tangential tractions
\begin{equation}\label{eq:BC4}
  (\delta_{ij}-n_in_j)(t_{j}-T_{j}
  )(\bm{r})=0,\qquad \bm{r}\in \partial V.
  \end{equation}
  The balance of normal tractions is not needed to determine the flow at given $f_l(t)$, but it can be used to calculate the active tractions responsible for the given deformations, if necessary. 

Just as for vesicles, the shape evolution of the droplet is described by a level set function and 
Eqs.(\ref{eq:BCinfinity}, \ref{eq:BCV2}, \ref{eq:BC3},\ref{eq:BC4}) determine a unique solution for the outer and inner flow fields.

  \section{Perturbation expansion}
\label{sec:perturbationExpansion}
  To set up a perturbation expansion in small deformations of a spherical particle, we introduce an order counting parameter $\varepsilon$, which multiplies the amplitudes $f_l$ and is set to 1 at the end of the calculations. It turns out that the expansion proceeds in powers of both deformation $f$ and its time derivative $\dot{f}$; therefore, the expansion requires $f$ and $\dot{f}$ to be of the same order of smallness. Up to the second order, the constraints of constant volume and constant surface take the form \cite{Seifert1999}
  \begin{align}
  \label{eq:constant_volume}
    f_0 & =-\varepsilon^2\sum_{l\geq 2}\frac{f_l^2}{2l+1} + \mathcal{O}(\varepsilon^3),\\
\label{eq:area-constraint}
  \Delta & =\varepsilon^2\sum_{l\geq 2}\frac{(l+2)(l-1)}{2(2l+1)}f_l^2 + \mathcal{O}(\epsilon^3).
\end{align}
The volume constraint Eq.(\ref{eq:constant_volume}) will be used to eliminate $f_0$ at $\mathcal{O}(\varepsilon^2)$. 
  
  For uniaxial and achiral systems, we expand the general external solution of Stokes equations\cite{Kree2023} in Legendre polynomials $P_l(\theta)$ and their derivatives $P_l'(\theta)=dP_l(\theta)/d\theta$:
  \begin{align}
  \label{eq:vrexp}
    v_r(r,\theta)&=\sum_{l\geq 1}\Big(-\frac{a_l}{r^{l+2}}+
                   \frac{b_l}{r^l}\Big)(l+1)P_l(\theta)-UP_1(\theta),\\
    \label{eq:vtexp}
     v_{\theta}(r,\theta)&=\sum_{l\geq 1}\Big(\frac{a_l}{r^{l+2}}-\frac{(l-2)b_l}{lr^l}\Big)P_l^{\prime}(\theta)-UP_1^{\prime}(\theta).
    \end{align}

    In a perturbative approach,
     the coefficients $a_l$ and $b_l$ of the flow field are expanded in powers of $\epsilon$, 
    \begin{align}\label{expansions}
      a_l&=\varepsilon a_l^{(1)}+\varepsilon^2 a_l^{(2)}+{\cal{O}}(\varepsilon^3),\\
      b_l&=\varepsilon b_l^{(1)}+\varepsilon^2 b_l^{(2)}+{\cal{O}}(\varepsilon^3)
    \end{align}
    giving rise to a corresponding expansion of the flow field:\quad $\bm{v}=\varepsilon\bm{v}^{(1)}+\varepsilon^2\bm{v}^{(2)}+{\cal{O}}(\varepsilon^3) $.\

    The perturbative analysis is carried out to the second order, since this is the minimum order that results in a non-zero swimming velocity $\bm{U}$. In the following, we present the results for deformable solid particles, vesicles, and droplets. All three use the above expansion Eq. (\ref{expansions}) in the appropriate boundary conditions. The computations are easiest for the solid particle; we therefore give the explicit steps of the calculation for this case in the next subsection and delegate some details of the calculations for vesicle and droplet to the appendices.

    \subsection{Deformable solid body}
    As a first step we expand the boundary condition Eq. (\ref{eq:BCsolid}) in powers of $f$,
\begin{equation}
\bm{v}(1+f,\theta)=\bm{v}(1,\theta)+f\partial_r\bm{v}(1,\theta) + \cdots=\dot{f}\bm{e}_r.
      \end{equation}
      To lowest order this equation reads
 \begin{equation}
v_r^{(1)}\bm{e}_r+v^{(1)}_{\theta}\bm{e}_{\theta}=\sum_{l\geq 2}\dot{f}_lP_l\bm{e}_r
\end{equation}
with the solution
         \begin{equation}
        a_l^{(1)}=\frac{l-2}{2(l+1)}\dot{f}_l, \quad
        b_l^{(1)}=\frac{l}{2(l+1)}\dot{f}_l
        \end{equation}
for $l\geq 2$.  The boundary condition in second order becomes
\begin{align}
  v_r^{(2)}+\sum_{l\geq 2}f_lP_l\partial_rv_r^{(1)}&= \dot{f}_0,\\
  v_{\theta}^{(2)}+\sum_{l\geq 2}f_lP_l\partial_rv_\theta^{(1)}&= 0.
       \end{align}
Substitution of the first order solution into the above equation yields       
\begin{align}
  \sum_{l\geq 1}  \Big(b_l^{(2)}-a_l^{(2)}\Big)&(l+1)P_l-U P_1\nonumber\\
  =& 2\sum_{l,m\geq 2} f_mP_m\dot{f}_lP_l+\dot{f}_0,\label{eq1}\\
\sum_{l\geq 1}  \Big(a_l^{(2)}-b_l^{(2)}(l-2)/l\Big)&P_l^{\prime}-U P_1^{\prime}\nonumber\\
  =&\sum_{l,m\geq 2}f_mP_m\dot{f}_lP_l^{\prime}\frac{l-2}{l+1}.\label{eq2}  
\end{align}
These equations determine the flow field as a function of the
prescribed deformations to $\mathcal{O}(\epsilon^2)$. In order to
obtain the coefficients $\{a_l^{(2)},b_l^{(2)}\}$ explicitly,
one has to project the above equations onto $P_l, P_l^{\prime}$.

Projection of Eq.\ref{eq1} onto $P_0$ yields
\begin{align}
  0=&2\sum_{l,m\geq 2} f_m\dot{f}_l\int_{-1}^{1}dx\,P_mP_l+\int_{-1}^{1}dx\,f_0
  \nonumber\\
  =&4\sum_{l}\frac{f_l\dot{f}_{l+1}}{2l+1}+2\dot{f}_0.
 \end{align} 
 Here we have used the orthogonality of the Legendre polynomials and the constraint of constant volume,
 $\dot{f}_0=-2\sum_{l\geq 2}f_l\dot{f}_l/(2l+1)$, which guarantees that the projection onto $P_0$ vanishes. We are mainly interested in the propulsion velocity of the particle, and hence project Eq.(\ref{eq1}) onto $P_1$ and Eq.(\ref{eq2}) onto $P_1^{\prime}=P_1^{1}$,
 the latter denoting the associated Legendre polymomial $P_1^{1}$.
The calculation requires integrals of 3 associated Legendre polynomials, which are given in Appendix \ref{appendix_integrals}. As a result of this procedure, we obtain 2 equations for
$ a_1^{(2)},b_1^{(2)}$ and $U$:

\begin{align}
\label{eq:secondorderSolid}
  2&b_1^{(2)}-2{a}_1^{(2)}-U=2\sum_{l\geq 2}\frac{2(l+1)}{(2l+1)(2l+3)}\partial_t(f_l f_{l+1}),\nonumber\\
2&(b_1^{(2)}-{a}_1^{(2)}-U)\nonumber\\ 
 & =\sum_{l}\frac{3}{(2l+1)(2l+3)}\big ((l^2-1)f_l\dot{f}_{l+1}-l(l-2)) f_{l+1}
     \dot{f}_{l}\big).\nonumber\\
 \end{align} 
The particle has to be
autonomous, that is, the total force has to vanish, which implies that the flow
has to decay faster than $1/r$ for large $r$ and hence $b_1^{(2)}=0$. 

The solution of this system is a linear combination of the inhomogeneities (right hand sides) and therefore consists of terms
proportional to $f_l\dot{f}_{l+1}$ and to $f_{l+1}\dot{f}_{l}$. This property also applies to the other two types of particles, so that we can always give the swimming velocity in the form

%The remaining 2 variables, $ a_1^{(2)}$ and $U$, are uniquely determined by the above equations. For $U$ we get
\begin{align}
\label{eq:Uparticle}
 % U&=-\sum_{l\geq 2}\frac{(l+1)^2f_l\dot{f}_{l+1}+(2+4l-l^2)
 %     f_{l+1}\dot{f}_{l}}{(2l+1)(2l+3)}\\
      U &= \sum_{l\geq 2}\big( R_l f_l\dot{f}_{l+1} + S_l  f_{l+1}\dot{f}_{l}\big).
%\label{eq:aparticle}
%  a_1^{(2)}&=\sum_{l\geq 2}\frac{(l^2+2l-6)f_l\dot{f}_{l+1}-(l-2)^2
%      f_{l+1}\dot{f}_{l}}{(2l+1)(2l+3)}.
\end{align}
For the solid body, the coefficients $R_l$ and $S_l$ are obtained from Eq.(\ref{eq:secondorderSolid}) and take on the form
\begin{align}
    R_l^{solid} &= -\frac{(l+1)^2}{(2l+1)(2l+3)},\\
    S_l^{solid} &= -\frac{2+4l-l^2}{(2l+1)(2l+3)}.
\end{align}

Note that the swimming velocity of deformable solid bodies is independent of the viscosity $\eta$, and therefore, it is a purely geometric quantity \cite{Shapere1}.
These results are not new, but have been derived previously, first by
Lighthill~\cite{Lighthill} and later corrected by Blake~\cite{Blake}. We have presented them here in detail to demonstrate our strategy, which we now apply to both the droplet and the vesicle.

\subsection{Vesicle}
For the vesicle, the calculation steps are nearly identical to those used for the solid particle; it is only the form of the boundary conditions that differs.
To first order, the boundary conditions (Eqs.\ref{eq:BCV1},\ref{eq:BCV2})  read
\begin{equation}
  v_r^{(1)}=\sum_{l\geq 2}\dot{f}_l P_l, \qquad \partial_r v_r^{(1)}=0,
\end{equation}
and the solution is
\begin{equation}
\label{eq:vesiclefirstorder}
  a_l^{(1)}=\frac{l\dot{f}_l}{2(l+1)},  \qquad
  b_l^{(1)}=\frac{(l+2)\dot{f}_l}{2(l+1)}.
\end{equation}
To second order we get
\begin{align}
  v_r^{(2)}&=\dot{f}_0+f^{\prime} v_{\theta}^{(1)},\\
  \sigma_{rr}^{(2)}=&2 f^{\prime}\sigma_{r\theta}^{(1)}-f\partial_r \sigma_{rr}^{(1)}.
\end{align}
Substituting the first-order solution from Eqs. (\ref{eq:vesiclefirstorder}, \ref{eq:vtexp}) and the results of Appendix \ref{appendix_stress}, yields
\begin{align}\label{eq:Vnonlin1}
 % \sum_{l\geq 1}\big(&(l+2)a_l^{(2)}-lb_l^{(2)} \big)(l+1)P_l
\sum_{l\geq 1}&\big(-a_l^{(2)}+b_l^{(2)} \big)(l+1)P_l-U\,P_1
                       =\nonumber\\
  &\sum_{l,m\geq 2}\frac{2f_m\dot{f}_l}{l(l+1)}P_m^1P_l^1
    -2\sum_{l\geq 2}\frac{2f_l\dot{f}_l}{l(l+1)},\\
  \label{eq:Vnonlin2}
%  \sum_{l\geq 1}&\big(-a_l^{(2)}+b_l^{(2)} \big)(l+1)P_l-U\,P_1
 \sum_{l\geq 1}\big(&(l+2)a_l^{(2)}-lb_l^{(2)} \big)(l+1)P_l=\nonumber\\
  &\sum_{l,m\geq 2}f_m\dot{f}_l(l+2)\big(l P_mP_l-
    \frac{1}{l(l+1)}P_m^1P_l^1\big). 
  \end{align}
Using both the constant volume and the constant area constraint, the projection of the above two equations onto $P_0$ can be shown to vanish.  The projections onto $P_1$ lead to two equations for $a_1^{(2)},b_1^{(2)}$ and $U$. We require the system to be force free, implying $b_1^{(2)}=0$, so that the resulting equations determine 
$a_1^{(2)}$ and $U$ in second order in the deformation. The swimming velocity takes the form of Eq.(\ref{eq:Uparticle})
with 
\begin{align}
    R_l^{ves} &= -\frac{l^3 + 4l^2 +10l +3}{(2l+1)(2l+3)},\\
    S_l^{ves} &= -\frac{(l+2)(l^2+4)}{(2l+1)(2l+3)}.
\end{align}
Just as for the solid particle, the swimming velocity $U$ does not depend on the viscosity $\eta$ .  The velocity $U$ differs from the result given in \cite{Misbah2013} by a total time derivative, which may arise from a different choice of reference point within the deformable vesicle.

\subsection{Droplet}
The analysis of the droplet requires both inner and outer solutions of the Stokes
equation. The outer solutions $\bm{v}$ are taken from 
Eqs.(\ref{eq:vrexp}, \ref{eq:vtexp}), the  inner solutions  $\bm{V}$ are given by
\begin{align}
\label{eq:Vrexp}
    V_r(r,\theta)&=\sum_{l\geq 1}\big( c_l r^{l-1}+
                   {d_l} r^{l+1}\big) lP_l(\theta)-UP_1,\\
\label{eq:Vtexp}
  V_{\theta}(r,\theta)&=\sum_{l\geq 1}\big(c_l r^{l-1}+\frac{(l+3)}{l+1}
                        {d_l}r^{l+1}\big)P_l^{\prime}(\theta)-UP_1^{\prime}.
\end{align}
The boundary conditions in first order read
\begin{align}
\label{eq:dropbc1}
 v_r^{(1)}&= V_r^{(1)}=\sum_{l\geq 2}\dot{f}_l P_l,\\
 \label{eq:dropbc2}
  v_{\theta}^{(1)}&= V_{\theta}^{(1)},\\
  \label{eq:dropbc3}
  \sigma_{\theta r}^{(1)}& = \Sigma_{\theta r}^{(1)}.                
\end{align}
The required stress components of the inner and outer flow are calculated in the Appendix \ref{appendix_stress}. 

  In second order the equations \ref{eq:BC3}, \ref{eq:BC4} and \ref{eq:BCV2} take on the form:
  \begin{align}
  \label{eq:secondOrderBCDrop}
   & v_r^{(2)}+f\partial_rv_r^{(1)}-f^{\prime}v_{\theta}^{(1)}=\dot{f}_0,\nonumber\\
   & V_r^{(2)}+f\partial_rV_r^{(1)}-f^{\prime}V_{\theta}^{(1)}=\dot{f}_0,\nonumber\\
   & v_{\theta}^{(2)}+f\partial_rv_{\theta}^{(1)}=
     V_{\theta}^{(2)}+f\partial_rV_{\theta}^{(1)}, \nonumber\\
   & \sigma_{\theta r}^{(2)}+f \partial_r\sigma_{\theta r}^{(1)}+f^{\prime}\Big(
     \sigma_{r r}^{(1)}-\sigma_{\theta \theta}^{(1)}\Big)=\nonumber\\
   &  \Sigma_{\theta r}^{(2)}+f \partial_r\Sigma_{\theta r}^{(1)}+f^{\prime}\Big(
    \Sigma_{r r}^{(1)}-\Sigma_{\theta \theta}^{(1)}\Big).
  \end{align}
We substitute the results of the first-order solution for the velocities and stress components in the above equations, which must be solved for the coefficients $a_1^{(2)},c_1^{(2)},\hat{d}_1^{(2)}$ and $U$. 
 The resulting $R_l$ and $S_l$ in Eq. \ref{eq:Uparticle} (and therefore also $U^{drop}$) are rational functions of the form 

%\begin{align}
%\label{eq:Rdrop}
%    R_{l}^{drop} & = \frac{r_{l,2}\lambda^2 + %r_{l,1} \lambda + r_{l,0}}{(\lambda + 1)(3\lambda %+2)}  \\
%\label{eq:Sdrop}
%    S_{l}^{,drop} & = \frac{s_{l,2}\lambda^2 + %s_{l,1} \lambda + s_{l,0}}{(\lambda + 1)(3\lambda %+2)}   
%\end{align}
\begin{align}
\label{eq:Rdrop}
    R_{l}^{drop} & = \frac{\sum_{n=0}^4 l^nr_l^{(n)}(\lambda)}{(3\lambda +2)(\lambda +1)(2l+1)(2l+3)^2},  \\
\label{eq:Sdrop}
    S_{l}^{,drop} & =  \frac{\sum_{n=0}^4 l^ns_l^{(n)}(\lambda) }{(3\lambda +2)(\lambda +1)(2l+1)^2(2l+3)}.  
\end{align}
%$  \dot{E}^{out}_{drop} & =\eta^+ \frac{e_2\lambda^2 + e_1\lambda + e_0}{(\lambda+1)^2}$
The second-order polynomials $r_{l}^{(n)}(\lambda)$ and $s_{l}^{(n)}(\lambda)$ can be found in Appendix \ref{app:droplet-second-order}.  
Unlike a solid particle or a vesicle, the swimming velocity of a droplet depends on the viscosity contrast $\lambda$. 
Note that the l-components of the velocity  $U$
%, as well as of the dissipation $\dot{E}^{out}_{drop}/\eta$ and  $\dot{E}^{in}_{drop}/(\lambda\eta)$ 
have well-defined limits for $\lambda \to 0$ and for $\lambda\to \infty$.

\subsection{Center of mass velocity}
The calculated swimming velocities $\bm{U}$ of the particles refer to the motion of the center of deformation in the laboratory frame.  We can transform them to center-of-mass velocities in the laboratory frame by adding to $\bm{U}$ the second order result of Eq. \ref{eq:vcm}, given by  
\begin{equation}
    \bm{v}_{cm}= \bm{e}_z\sum_l \frac{9(l+1)}{(2l+1)(2l+3)} \frac{d}{dt}(f_lf_{l+1}). 
\end{equation}

\section{Energy dissipation}
\label{sec:efficiency}
The dissipated power in an incompressible Stokes flow in a volume $V$ is given by
\begin{equation}
  {\dot{E}}=\int_V d^3x\; \sigma_{ij}\partial_jv_i
  =\int_{\partial V}d^2x \; n_j\sigma^{visc}_{ij}v_i.
\end{equation}
Pressure does not contribute to this expression, so $\sigma_{ij}$ can be replaced by viscous stress $\sigma^{visc}_{ij}$. 
In spherical coordinates the integral takes on the form
\begin{equation}
\label{eq:dissipation_integral}
  {\dot{E}}=\int_{\partial V}d^2x \; \Big(\sigma_{rr}^{visc}v_r+\sigma_{r\theta}^{visc}v_\theta\Big).
\end{equation}
To evaluate the leading order, we need the flow field only in ${\cal{O}}(\epsilon)$ and the integration is over the reference sphere.  For both outer and inner flow fields, the two terms are evaluated in Appendix \ref{app:dissipation}. 

Our model does not take into account interior dissipation of the solid particle. The power due to the ambient flow, 
\begin{equation}
\label{eq:dissipate_particle}
  \dot{E}_{sb}=-16\pi\eta\sum_{l\geq 2}\frac{\dot{f}_l^2}{2l+1},
\end{equation}
is therefore only a lower bound on the total dissipation. 

For the vesicle, the  dissipated power due to the outer flow takes on the form
\begin{equation}
\label{eq:dissipate_vesicle_out}
  \dot{E}^{out}_{ves}=-4\pi\eta\sum_{l\geq 2}\frac{l+2}{(2l+1)l(l+1)}\dot{f}_l^2,
\end{equation}
and the inner flow contributes
\begin{equation}
\label{eq:dissipate_vesicle_in}
  \dot{E}^{in}_{ves}=-4\pi\lambda\eta\sum_{l\geq 2}\frac{l-1}{(2l+1)l(l+1)}\dot{f}_l^2.
\end{equation}

The $l$-components of outer and inner dissipation of the droplet, 
\begin{equation}
\label{eq:dissipate_drop}
    \dot{E}^{in/out}_{drop}= -\eta \sum_l D^{in/out}_l(\lambda)\dot{f}_l^2
\end{equation}
are given explicitly in appendix \ref{app:dissipation}.

\section{Comparisons}
\label{sec:comparisons}
We will now use the general results of Eq. (\ref{eq:Uparticle}) and
Eqs. (\ref{eq:dissipate_particle}-\ref{eq:dissipate_drop}) to discuss
and compare the swimming velocities, the dissipation and the energetic
efficiencies of the three types of particles, when they are driven by
periodic deformations with period $T$. 

%\subsection{Swimming velocities due to periodic deformations}
As a result of Sect. \ref{sec:perturbationExpansion}, the average
swimming velocities
\begin{equation}
 \overline{U}=(1/T)\int_0^T dt U(t) =\sum_l(1/T)(R_l-S_l)\int_0^Tdt\dot{f}_{l+1}f_l
\end{equation}
in second order can be represented as a sum
$\overline{U}=\sum_l\overline{U_l}$ of terms $\overline{U}_l$, which
contain the time averages of $\Phi_l(t)=\dot{f}_{l+1}f_l $. Note that
total time derivatives, such as $\bm{v}_{cm}$, do not change
$\overline{U}$. Therefore, the average swimming velocity is a physical
quantity that is independent of the chosen body fixed reference frame.
Differences in swimming velocities between the particle types arise
solely from the numerical values of $R_l-S_l$. These values are
constants for the solid and the vesicle (see Table 1), whereas for
droplets they decrease with increasing viscosity contrast $\lambda$,
such that the bubble ($\lambda=0$) is the fastest.

In the following, we discuss two special examples in detail to
illustrate our general results: (1) periodic driving with just two
adjacent amplitudes $f_l{f}_{l+1}$ and (2) simple harmonic motion with 3
adjacent amplitudes, which allow for optimization of strokes either
with respect to speed or efficiency. Whenever necessary for de-dimensionalization, 
we use $T/2\pi$ as a unit of time in the following subsections.

\subsection{Swimming velocity ((l, l+1)-strokes)}

\begin{figure}[htb!]
\stackinset{l}{120pt}{t}{10pt}{\includegraphics[width=0.2\textwidth]{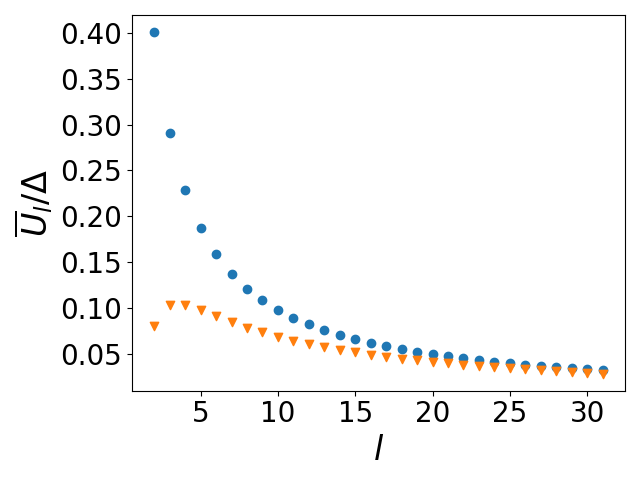}}{\includegraphics[width=0.45\textwidth]{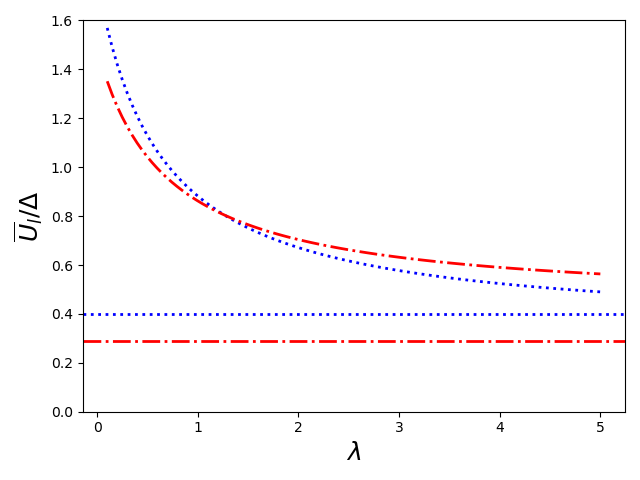}}  
  \caption{Average swimming speed $\overline{U}_l$ (divided by excess area $\Delta$), generated from $(f_l,f_{l+1})$-pairs, of a droplet (curved line) and a vesicle (horizontal lines) versus viscosity contrast $\lambda$ for two pairs. Dotted lines correspond to $l=2,3$ and dash-dotted lines to $l=3,4$. The inset shows  the $\lambda$-independent $\overline{U}_l/\Delta$ for a vesicle and a solid body versus $l$.}  
  \label{fig:R-SvsLambda}
\end{figure}

In general, the time average $\overline{\Phi}_l$ depends on the
detailed functional form of the deformation amplitudes $f_l(t)$ and
$f_{l+1}(t)$. For droplets and deformable solids these amplitudes are
unconstrained, but for vesicles they have to obey the area constraint
Eq. (\ref{eq:area-constraint}). If only one $(l,l+1)$-pair of
amplitudes is nonvanishing, this restricts the manifold of possible
deformations to an ellipse in the $(f_l, f_{l+1})$-plane, and
$T\overline{\Phi}_l$ becomes the area of the ellipse \cite{Lighthill},
\cite{Misbah2013}, that is
\begin{equation}
    T\overline{\Phi}_l=\int_0^T \dot{f}_{l+1}f_l = \oint f_l df_{l+1}= \pi \Delta\sqrt{\beta_l\beta_{l+1}}, 
    \label{eq:ellipse_area}
\end{equation}
with $\beta_l= 2(2l+1)/[(l+2)(l-1)] $. 
This shows that $\overline{U}$ does not depend on the time course of the deformation amplitudes. 
 For $l=2$ we recover the result $T\overline{U} = - 3\pi \Delta/\sqrt{14}$ given in \cite{Misbah2013}. \\
 In Fig. \ref{fig:R-SvsLambda}, we show the speed $|\overline{U}_l|$ of the droplet
 %which shows the rescaled
 %speed $W_l=|R_l-S_l|= |\overline{U}_l/\overline{\Phi}_l|$
 for two
$(l,l+1)$-pairs as a function of $\lambda$ in comparison to those of a vesicle. The $l$-dependence of the swimming speeds
of a vesicle and a solid body are shown in the inset of Fig. \ref{fig:R-SvsLambda}

In general, all $|\overline{U}_l|$ of droplets decrease with increasing $\lambda$. For small $\lambda$, the swimming speeds also decrease with increasing $l$, while for $\lambda \gtrsim  1.3$, they begin to increase with increasing $l$. 
Note also that unlike a vesicle the values of $|\overline{U}_l|$ for a solid body increase for $l=2,3$  before they start to decrease significantly from $l=5$.

\subsection{Dissipation}

\begin{figure}[htb!]
  \includegraphics[width=0.45\textwidth]{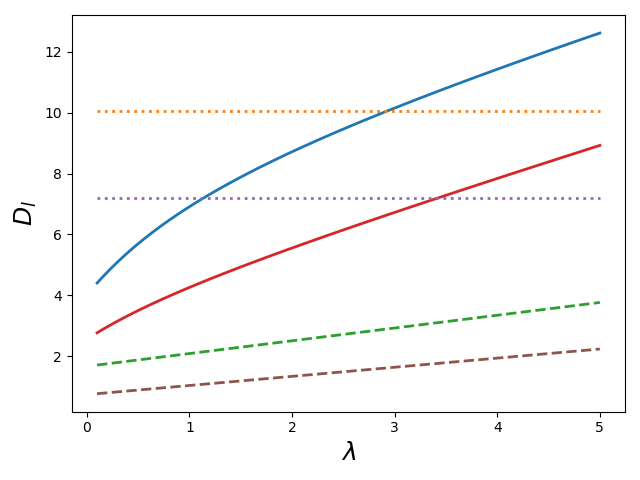}
  \caption{Dissipation coefficients $D_l$ vs. viscosity contrast $\lambda$. Solid lines: drop, dashed lines: vesicle, dotted lines: solid particle. Upper lines refer to $l=2$, lower lines to $l=3$}
  \label{fig:DlwithVesicle}
\end{figure}

In leading order, each deformation amplitude $f_l(t)$ contributes to
the dissipated energy -- even if only a single driving amplitude is
present and the particle is not propelled. We make use of Eqs. (\ref{eq:dissipate_particle}-\ref{eq:dissipate_drop}) to compute the energy $\overline{E}_l = -\eta T \overline{\dot{f}_l^2} (D_l^{out} +
D_l^{in})$ that is dissipated in a period $T$. For our model of a deformable solid,
$D^{in}=0$, so that $\overline{E}_l$ is independent of $\lambda$. For a vesicle it is linear in the viscosity contrast
$\lambda$, and for a drop it is the quotient of two second order
polynomials in $\lambda$, multiplied by a factor $\lambda$.  Due to
the absence of internal dissipation, the total dissipated energy of
the solid particle will always be the smallest for large $\lambda$. In
Fig. \ref{fig:DlwithVesicle} we compare $D_l=D_l^{out} + D_l^{in}$ of
vesicle and drop (with the same internal and external viscosities) to
that of the solid particle for $l=2$ and $3$.

For all particle types, the dissipation coefficients $D_l$ become smaller with increasing $l$.
The dissipated energy is always lower for vesicles than for droplets. For droplets, $D_2$ ($D_3$) reaches the value of the solid particle at $\lambda\approx 2.9$ ($\lambda\approx 3.4$). 
For vesicles the crossover appears at larger values
($\lambda\approx 20$ for $l=2$ and $22$ for $l=3$). 

\subsection{Simple harmonic deformations with area constraint}
For deformations consisting of more than two adjacent amplitudes $f_l, f_{l+1}$, the average speed $\overline{U}$ depends on the time course of the constrained swim strokes. In the following, we restrict the time dependence to simple harmonic motion, for which a more detailed discussion is possible.  The deformations consist of a sum of terms $f_l(t)P_l(\cos\theta)$ with
\begin{equation}
\label{eq:simpleHarmonicDeform}
    f_l(t)=F_l \left(e^{i\alpha_l + i\omega t} + e^{-i\alpha_l - i\omega t}\right).
\end{equation}
Such strokes lead to particle velocities
\begin{equation}
\begin{split}
      U=\sum_l 2\omega F_lF_{l+1}\Big((R_l-S_l)\sin(\alpha_l-\alpha_{l+1})\\
      - (R_l+S_l) \sin(2\omega t + \alpha_l+\alpha_{l+1})\Big),
\end{split}
\label{eq:Ugeneral}
\end{equation}
when inserted in Eq.(\ref{eq:Uparticle}).   Note that one of the phases $\alpha_l$ can be arbitrarily chosen by appropriately adjusting the origin of the time axis. The time-independent term is the mean swimming velocity $\overline{U}$. Fourier components with different frequencies $\omega_l$ for different $l$-components do not contribute to the time average $\overline{U}$, so we can consider a single $\omega$ without loss of generality. The choice of time unit $2\pi T$ corresponds to $\omega=1$ for the harmonic deformations.  

In the following, we discuss the simple non-trivial case of three $l$ components, $l=2,3$ and 4. A swim stroke is then characterized by the parameters $F_2,F_3,F_4, \alpha_2,\alpha_3$ ($\alpha_4$ is set to zero by choosing a suitable initial time). These five parameters must satisfy three equations, derived by inserting Eq.(\ref{eq:simpleHarmonicDeform}) into Eq.(\ref{eq:area-constraint}) and separating the time independent and time dependent parts. The resulting two-parameter manifold can be conveniently described by the $l=2$ parameters $F_2,\alpha_2$, as detailed in Appendix \ref{app:swim-strokes}.  For each pair $(F_2, \alpha_2)$ there are two swim strokes, corresponding to motion in either the $(+ z)$-- or the $(-z)$--direction with identical speed. Consistent strokes can be found for every $\alpha_2$, while the range of $F_2$ is limited to $0 \leq F_2 \leq F_{2,max}$ ( in Appendix \ref{app:swim-strokes} we show that $F_{2, max}=\sqrt{5\Delta/8}$ ).    
Now we can compare the trajectories, the swimming velocities and the dissipated power of a solid body, a vesicle, and a drop (all three of equal volume), which execute all possible area-conserving swim strokes composed of $l=2,3$ and $4$ components.

\subsection{Swimming velocities with area constraint}
\label{subsec:swimvelocity}
\begin{figure}[htb!]
  \includegraphics[width=0.45\textwidth]{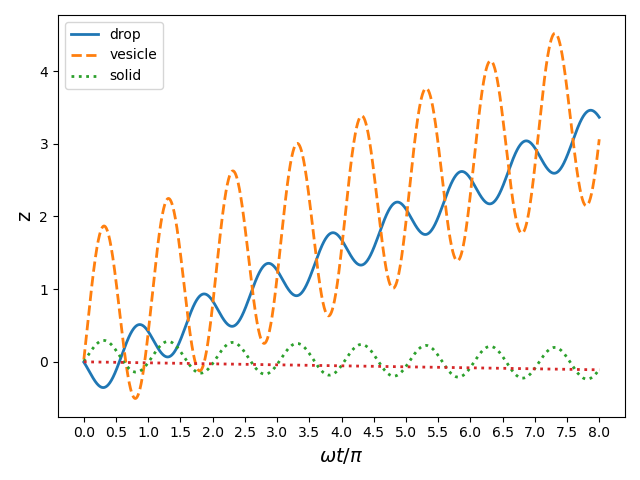}
  \caption{Example of trajectories: $\alpha_2/2\pi=0.952, F_2/F_{2,max}=0.78$, $\lambda=1$. solid line: droplet, dashed line:vesicle, dotted line:solid particle. The dotted straight line $\overline{U}t$ highlights the motion of the solid particle  in the ($-z$)-direction}
  \label{fig:trajec}
\end{figure}

For a comparison of average swimming velocities with constrained $(l=2,3,4)$-swim strokes, we use the entries of Table (\ref{tab:RS}).   An example of trajectories for a fixed stroke is shown in Fig.\ref{fig:trajec}. It illustrates that the speeds and oscillation amplitudes vary between particles, and it is even possible for the same swim stroke to propel different particles in opposite directions.  

% In the simplest situation $F_4=0$, and the
% Eq. (\ref{eq:Ugeneral}) results in a mean velocity of the form
% \begin{equation}
%     \overline{U} =  \pm2\omega F_2F_3(R_2-S_2)=\pm\frac{7\omega\Delta}{16}(R_2-S_2).
% \end{equation}
% From now on, we will concentrate on the $(+)$--branch of solutions.
% The speed of the droplet decreases monotonically with $\lambda$ (from $|R_2-S_2|\approx 2.599$ at $\lambda=0$ to $|R_2-S_2| \approx 1.177$ at $\lambda\to \infty$), but is always greater than the swimming speed of the particle ($|R_2-S_2| =3/35 $) and the vesicle ($|R_2-S_2|= 3/7$). In a race between the three particles, the finishing order stays the same: The droplet is the fastest, followed by the vesicle in second place, and the solid particle is the last. 

\begin{table}   
    \begin{tabular}{|c|c|c|c|}
    \hline
        & solid & vesicle & droplet\\
    \hline
        $R_2$ & -9/35 & -47/35 & $-3(219\lambda^2 +545\lambda +286)/(245\mu)$ \\
        $S_2$ & $-6/35$ & $-32/35$ & $6(-47\lambda^2 -61\lambda + 8)/(175\mu)$\\
        $R_3$ & $-16/63$ & $-96/63$ & $-8(23\lambda^2 +57\lambda + 31)/(63\mu)$\\
        $S_3$ & $-5/63$ & $-65/63$ &  $(-103\lambda^2 + 13\lambda + 160)/(147\mu)$\\ 
        \hline
    \end{tabular}
    \caption{The coefficients $R_l$ and $S_l$, which determine the swimming velocities (see Eq. (\ref{eq:Uparticle})) for $l=2$ and $3$. These coefficients are used in Sect \ref{subsec:swimvelocity}. In the third column, $\mu=(\lambda + 1)(3\lambda + 2)$.} 
    \label{tab:RS}
\end{table}

\begin{figure}[htb!]
    \centering
    \includegraphics[width=0.45\textwidth]{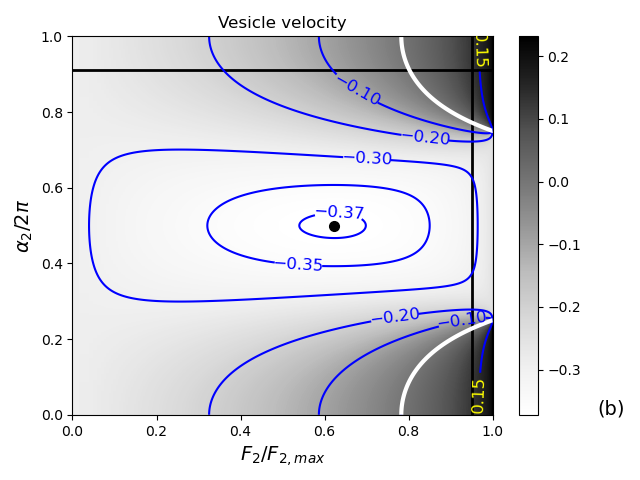}\\
    \includegraphics[width=0.45\textwidth]{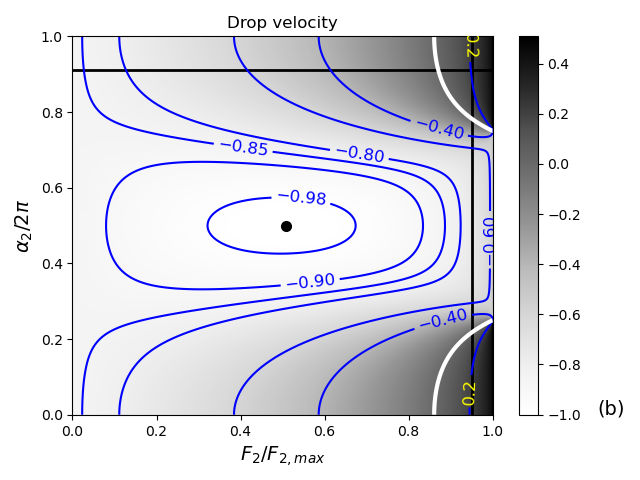}\\
     \includegraphics[width=0.45\textwidth]{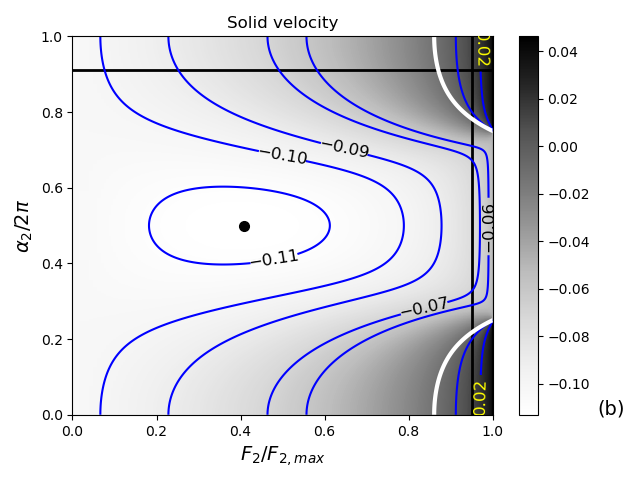}
    \caption{Contour plots of swimming velocity $\overline{U}$ over the entire parameter space $(F_2, \alpha_2)$ at $ \lambda=1$: (a) vesicle: The  maximum vesicle speed (black dot) is $0.372$ at $F_2/F_{2,max}=0.6224$, (b) droplet: The maximum drop speed is $1.001$ at $F_2/F_{2,max}=0.5081$,  (c) solid particle: The maximum particle speed is $0.113$ at $F_2/F_{2,max}=0.4089$,. The slices marked by (black) solid lines are shown in Fig. \ref{fig:UF2andAlpha}} %(d) at $\lambda=5$: maximum drop speed $=0.601$ at $F_2/F_{2,max}=0.536$. }
    \label{fig:Ucontour}
\end{figure}

\begin{figure}[htbp!]
    \centering
    \includegraphics[width=0.45\textwidth]{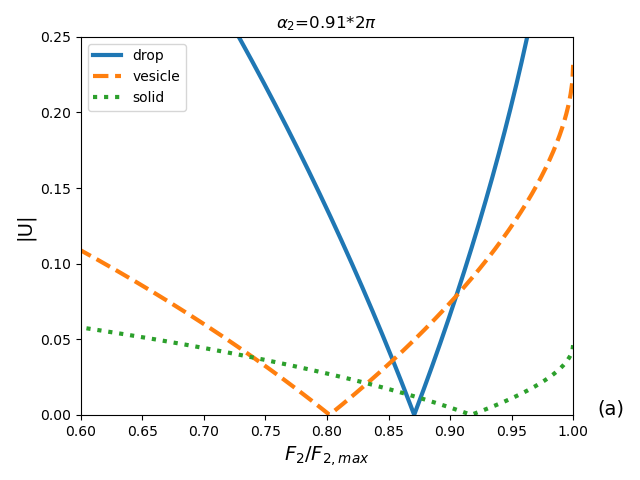}\\
    \includegraphics[width=0.45\textwidth]{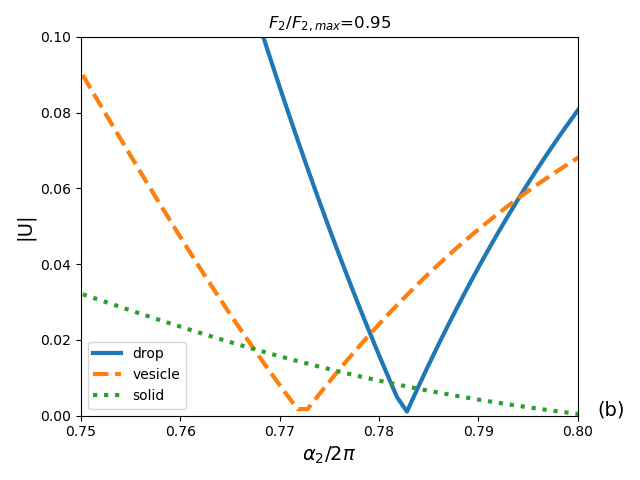}
    \caption{Swimming speed along the solid lines shown in Fig. \ref{fig:Ucontour} near the zeros of $|U|$. (a) at fixed $\alpha_2/2\pi=0.91$, (b) at fixed $F_2/F_{2,max}=0.95$.}
    \label{fig:UF2andAlpha}
\end{figure}

The results for the average swimming velocities are summarized in Fig. \ref{fig:Ucontour}.
The parameter space of all possible constrained swim strokes lies in the ($F_2, \alpha_2$)-plane.  In the contour plots, the maxima are marked with black dots. Slices along parts of the black lines are shown in Fig. \ref{fig:UF2andAlpha}. The white line indicates zero propulsion velocity.  

The general form of the dependence of $\overline{U}$ on the parameters $F_2$ and $\alpha_2$ is the same for all three types of particles and is determined by Eq. (\ref{eq:Ugeneral}). The maxima of $\overline{U}$ are always at $\alpha_2=\pi$ and the patterns  are symmetric with respect to the line $\alpha_2=\pi$.  The strokes leading to maximum swimming speed differ for the three types of particles, the $F_2$ value being largest for vesicles and smallest for solid particles.
Note that the swimming velocity reverses its direction across the white line. This line is thus a one-dimensional manifold of swim strokes, which do not propel the particle at all.

If each active swimmer is free to select the swim stroke that maximizes its velocity, the droplet wins for all $\lambda$, followed by the vesicle and the solid particle in the third position. Note that the maximum speed of the solid particle is a factor $\approx 10$ smaller than that of the drop.  However, if the three swimmers perform the same stroke, their order of ranking can change with the parameters.  This can be seen in Fig. \ref{fig:UF2andAlpha}, which shows the swimming speed $|U|$ versus $F_2/F_{2, max}$ (a) and versus $\alpha_2$ (b) for $\lambda=1$ along the cuts shown in Fig.\ref{fig:Ucontour} in the vicinity of the zeros of $\overline{U}$. At each crossing of two lines in the graph, a pair of particles changes their order in the speed ranking.   

\begin{figure}[htbp!]
    \centering
    \includegraphics[width=0.45\textwidth]{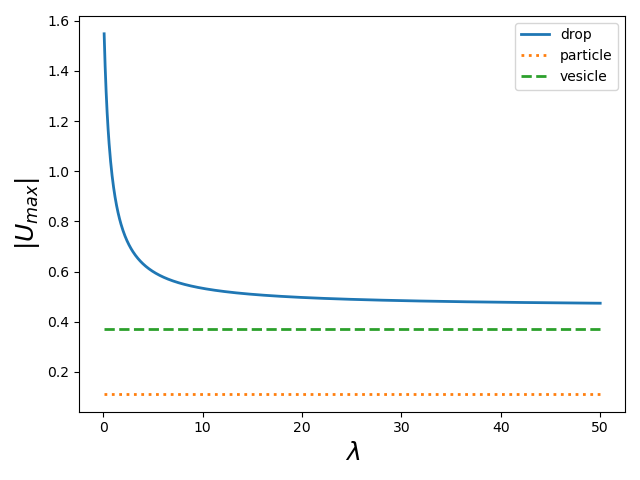}\\
    \caption{ Maximum speed vs. $\lambda$. The ranking (drop fastest, followed by vesicle, followed by solid particle) remains for all values of $\lambda$. }
    \label{fig:MaxUvsLambda}
\end{figure}

If $\lambda$ is increased, the average speed of a droplet will decrease. Nevertheless,  the maximum speed always remains higher than the maximum speed of the vesicle and the solid particle, as shown in Fig. \ref{fig:MaxUvsLambda}. 

\subsection{Lighthill efficiency with area constraint}
A given swim stroke, parametrized by $(F_2, \alpha_2)$, propels different particles with different velocities, but also needs different amounts of dissipated power. A frequently used measure of the energetic efficiency of the particle is due to Lighthill  \cite{Lighthill}. It is defined as the ratio of the power dissipated by a solid sphere (with the same volume as the particle) moving at the particle's mean swimming velocity $\overline{U}$ to the power dissipated by the particle itself, that is, 
\begin{equation}
    \epsilon = \frac{6\pi \eta \overline{U}^2}{\overline{\dot{E}}}. 
\end{equation}

\begin{figure}[htbp!]
    \centering
    \includegraphics[width=0.45\textwidth]{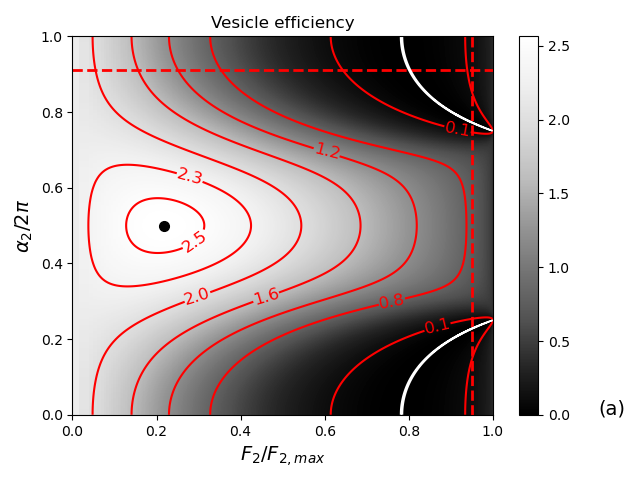}\\
    \includegraphics[width=0.45\textwidth]{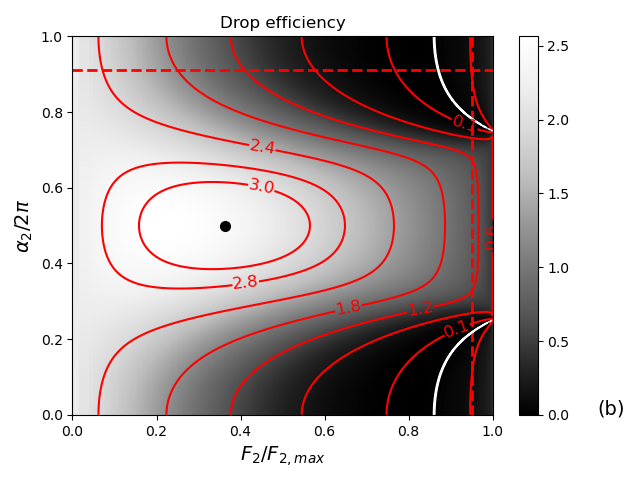}\\
     \includegraphics[width=0.45\textwidth]{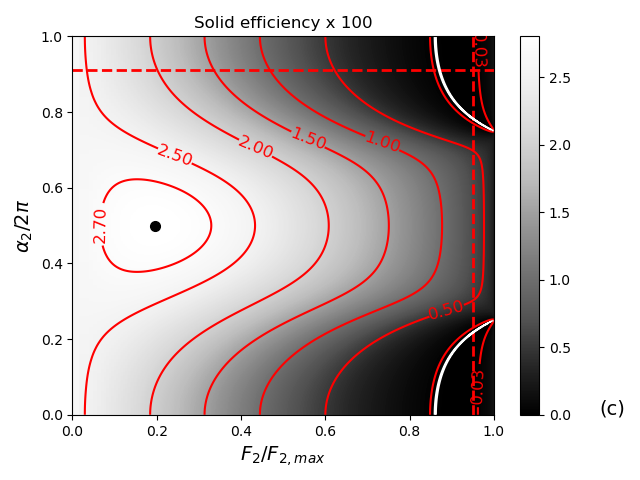}\\
    \caption{Contour plots of Lighthill efficiency over the entire parameter space $(F_2, \alpha_2)$ at $\lambda=1$ for (a) a vesicle, (b) a droplet, and (c) a solid. Dissipation from the ambient fluid and the interior (none in case of the solid) is taken into account. The maximum efficiency of a vesicle is $2.565$ at $F_2/F_{2,max}=0.217$, the maximum  efficiency of a droplet is $3.200$ at  $F_2/F_{2,max}=0.364$ and the maximum  efficiency of a solid is $0.0280$ at  $F_2/F_{2,max}=0.195$}
    \label{fig:EffDropandVesicle}
\end{figure}

\begin{figure}
\stackinset{l}{35pt}{t}{20pt}{\includegraphics[width=0.22\textwidth]{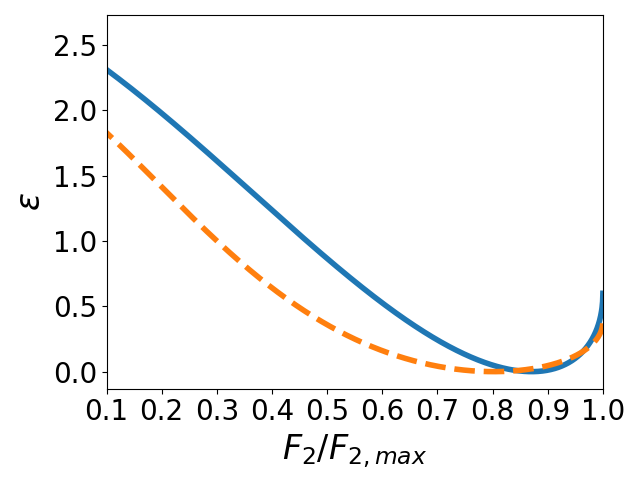}}{\includegraphics[width=0.45\textwidth]{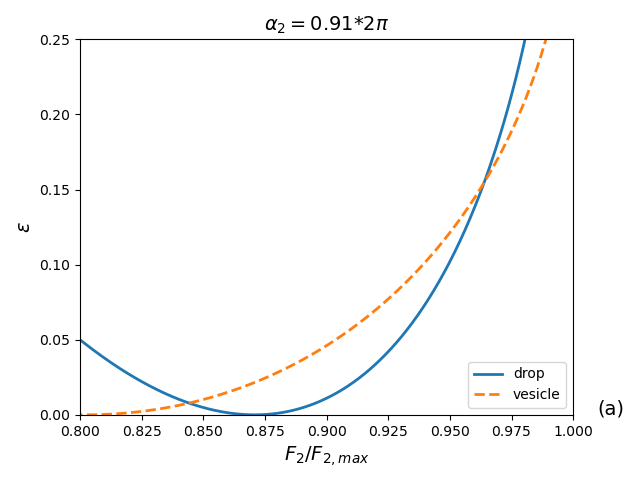}}
\stackinset{l}{55pt}{t}{25pt}{\includegraphics[width=0.22\textwidth]{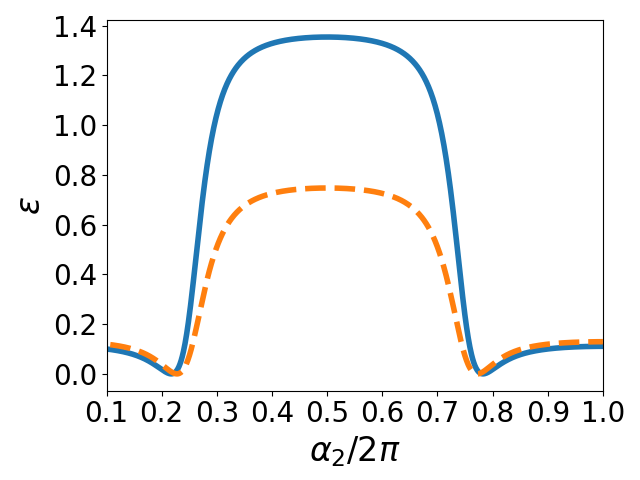}}{\includegraphics[width=0.45\textwidth]{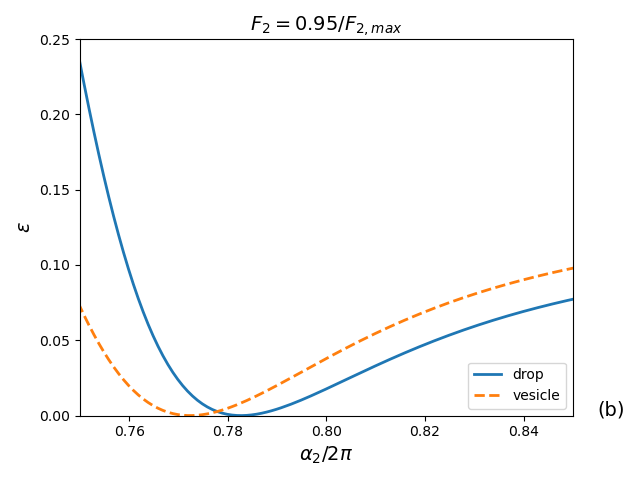}}
\caption{Lighthill efficiencies along the dashed lines of Fig \ref{fig:EffDropandVesicle} near the zeros of $\epsilon$.  (a) at fixed $\alpha_2/2\pi=0.91$, (b) at fixed $F_2/F_{2,max}=0.95$. The insets show the full range of $\epsilon$}
\label{fig:epscut}
\end{figure}

In Fig. \ref{fig:EffDropandVesicle}, we show contour plots of the Lighthill efficiencies for  a vesicle, a droplet and a solid particle at $\lambda=1$.  
The lines of vanishing $\overline{U}$ in Fig. \ref{fig:Ucontour} show up here as lines of zero efficiency.  In the vicinity of these zeros, the efficiencies change their order as shown in Fig. \ref{fig:epscut}  for a drop and a vesicle. The efficiency of the solid particle does not vanish at the zeros of drop and vesicle but is too small to be visible on the scale of the plots.  However, it is typically two orders of magnitude smaller than the efficiencies of droplets and vesicles. The small $\epsilon$ can already be inferred from the Figs \ref{fig:DlwithVesicle} and \ref{fig:Ucontour}, which show that the maximum speeds of the drop and the solid particle differ by a factor of $\approx 10$, while their dissipation coefficients $D_l$ differ only by factors between $1$ and $\approx 2$.    

\begin{figure}[htbp!]
    \centering
    \includegraphics[width=0.45\textwidth]{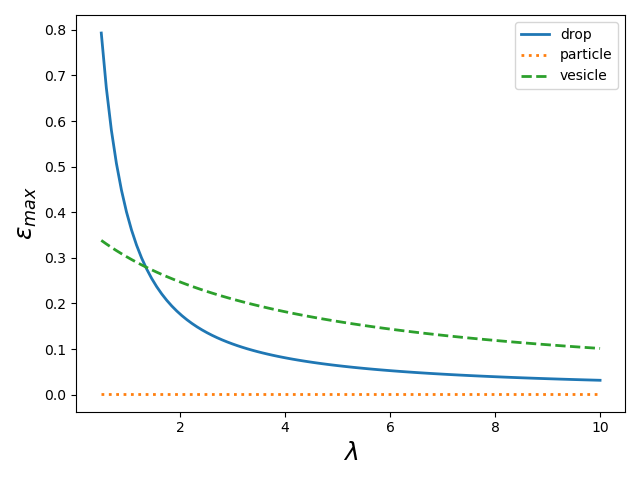}
    \caption{Total maximum efficiency vs. $\lambda$.}
    \label{fig:MaxvsLambda}
\end{figure}

 For small internal viscosities, $\epsilon_{max}$ of the droplet is higher than that of the vesicle, but it decreases with increasing $\lambda$ as shown in  In Fig. \ref{fig:MaxvsLambda}.   When $\lambda \approx 1.35$  the maximum efficiencies become equal, and with more viscous internal fluids the vesicle is the most efficient, although it moves slower than the droplet (as can be seen from Fig. \ref{fig:MaxUvsLambda}). Consequently, optimizing for speed and optimizing for efficiency yield different outcomes.

\section{Summary}
\label{sec:summary}
In this work, we have presented a unified discussion of  swimming velocities, dissipation and efficiencies of three types of near-spherical  amoeboid microswimmers, driven by axially symmetric achiral swim strokes: a solid deformable body, a vesicle with incompressible fluid membrane, and a droplet. Our approach uses (scalar) spherical harmonics to represent surface  deformations and a system of general solutions of the Stokes equation based on vector spherical harmonics. The swimming velocity and the dissipation have been calculated to second order of small deformation amplitudes and their time derivatives. The restriction to axial symmetry is sufficient to compare translational motion between the three types of swimmer. The minimal models describe the types of swimmers by appropriate boundary conditions. For solids and vesicles, these conditions couple the interior material to the surrounding Newtonian fluid only via surface deformations, whereas for a droplet, the interior and the ambient flows are additionally coupled by the condition of continuity of tangential stress at the interface.  As a consequence, the velocities  of solids and vesicles do not depend on the viscosity ratio $\lambda$, in contrast to droplets.  

To compare all three types of microswimmer, the swim strokes have to obey both the constraints of volume- and of surface-incompressibility. First, we studied swim strokes, which consist of pairs $f_l, f_{l+1}$  of deformations, varying periodically in time. In this case the average swimming velocity does not depend on the detailed time course of deformations. In general, all $|\overline{U}_l|$ of droplets resulting from  $f_l, f_{l+1}$   decrease with increasing $\lambda$. For small $\lambda$, the swimming speeds also decrease with increasing $l$, while for $\lambda \gtrsim  1.3$, they begin to increase with increasing $l$. Note also that unlike a vesicle the values of $|\overline{U}_l|$ for a solid body increase for $l=2,3$  before they start to decrease significantly from $l=5$. In a next step, we have constructed a two-dimensional manifold of simple harmonic strokes, which contains $l$ modes up to $l=4$ and can be parametrized by the $l=2$ parameters $F_2$ and $\alpha_2$. This allows us to present the results for the swimming velocities $U$ and the efficiencies $\epsilon$ in a clear and complete form as contour plots. 

From these plots, the results of races between the three microswimmers are easily obtained. 
When each swimmer can choose the stroke that maximizes its speed, the droplet always comes first regardless of $\lambda$, although its velocity decreases as $\lambda$ increases. The vesicle comes second, while the solid particle comes third. However, if the three swimmers perform the same stroke, their order of ranking can change with the parameters.  

 Optimizing the Lighthill efficiency and optimizing the swimming velocity result in different optimal swim strokes and rankings.
 The maximum total efficiency (based on the dissipation in the interior and surrounding flow) of a droplet is greater than that of a vesicle only if the dissipation ratio $\lambda$ is small. Beyond $\lambda \approx 1.35$ the maximum efficiency of the droplet falls below that of the vesicle. In all cases, the efficiency of the solid is typically two orders of magnitude smaller than that of vesicles and droplets, except in the vicinity of zeros of the efficiency. The efficiency of the solid particle will drop further if internal viscosity mechanisms are included in the model of the solid.  

 Our work can be extended in several directions. First, the restriction to axially symmetric swim strokes can be lifted by using deformation amplitudes $f(t,\theta, \phi)$, which depend on both polar angle $\theta$ and azimuthal angle $\phi$, and expanding $f$ in spherical harmonics. The calculational strategies presented above can be used if the set of vector spherical harmonics is enlarged to include rotational motion \cite{Kree2022}.  Furthermore, it is possible to study deformations other than radial (such as normal or tangential) by starting from appropriate variants of of Eq. (\ref{eq:paramdeformation}).
 Second, the complete internal and ambient flow can be calculated, by projecting the boundary conditions onto higher $l$-components. Third, surface or body forces (for example Marangoni forces), which generate the deformations $f$, can be included. This makes the model more difficult because the relation between forces and deformations becomes a differential equation and introduces new relaxation time scales. For vesicles with well-separated relaxation time scales, an approach to this problem can be found in reference \cite{Misbah2013}. Finally, the internal materials of the particles may be changed to complex (for example visco-elastic) fluids or elastic solids.

%%%%%%%%%%%%%%%%%%%%%%%%%%%%%%%%%%%%%%%%%
%Appendices
%%%%%%%%%%%%%%%%%%%%%%%%%%%%%%%%%%%%%%%%
\clearpage
\appendix

\section{First-order solution for the droplet}\label{append:droplet_first_order}
We insert the expansions introduced in Eqs. (\ref{eq:vrexp}, \ref{eq:vtexp}, \ref{eq:Vrexp}, \ref{eq:Vtexp}) in the first-order boundary conditions Eqs.(\ref{eq:dropbc1}, \ref{eq:dropbc2}, \ref{eq:dropbc3}), use Eq.(\ref{stress}) for the stress,  and project on fixed $l$ components. The resulting $4\times 4$ linear system takes on the form 
\begin{align}
  -(l+1)(a_l^{(1)}-b_l^{(1)}) & = \dot{f}_l, \label{linear_droplet1}\\
  l( c_l^{(1)}+d_l^{(1)})&= \dot{f}_l,\label{linear_droplet2}\\
  a_l^{(1)}-\frac{l-2}{l}b_l^{(1)}&=c_l^{(1)}+\frac{l+3}{l+1}d_l^{(1)},
                            \label{linear_droplet3}\\
  \frac{(l^2-1)}{l}b_l^{(1)}-a_l^{(1)}(l+2)=\nonumber\\ &\lambda\Big(
(l-1)c_l^{(1)} +\frac{l(l+2)}{l+1}d_l^{(1)} \Big)\label{linear_droplet4}.
  \end{align}
The four equations are easily reduced to two, because interior and exterior of the droplet are decoupled in the first two of them, implying
\begin{equation}
\label{eq:firstorderad}
b_l^{(1)}=a_l^{(1)}+\frac{\dot{f}_l}{l+1} \qquad \text{and}\qquad
 d_l^{(1)}=-c_l^{(1)}+\frac{\dot{f}_l}{l}  
\end{equation}
Substituting these expressions into Eq.(\ref{linear_droplet3}) yields the following equation:
\begin{equation}
 (l+1)a_l^{(1)}+l c_l^{(1)}=\frac{(2l+1)}{2}\dot{f}_l.
\end{equation}

Together with Eq.(\ref{linear_droplet4}) this fixes the coefficients $c_l^{(1)}$ and $a_l^{(1)}$:
\begin{align}
\label{eq:firstorderc}
  (\lambda+1)&lc_l^{(1)}=\frac{\dot{f}_l}{2(2l+1)}
  \big(2l^2+4l+3+2\lambda l(l+2) \big),\\
  \label{eq:firstordera}
(\lambda+1)&(l+1)a_l^{(1)}=\frac{\dot{f}_l}{2(2l+1)}
 \big(2(l^2-1)+\lambda(2l^2+1)\big).
\end{align}
The remaining coefficients $b_l^{(1)}$ and $d_l^{(1)}$ are obtained from Eq. (\ref{eq:firstorderad}).

  \section{Stress components}\label{appendix_stress}  
    The first-order stress components for the exterior flow, which are used in the boundary conditions of the vesicle and the droplet are calculated from 
    \begin{align}\label{stress}
      &\frac{1}{\eta}\sigma_{\theta r}=\frac{1}{r}\partial_{\theta}v_r +\partial_r v_{\theta}-\frac{1}{r}v_{\theta},\nonumber\\
      &\sigma_{\theta \theta}=2\eta(\frac{1}{r}v_r +\frac{1}{r}\partial_{\theta} v_{\theta})-p,\nonumber\\
       &\sigma_{r r}=2\eta\partial_{r}v_r-p.
      \end{align}
Substituting the flow velocity from Eqs.(\ref{eq:vrexp}, \ref{eq:vtexp}), we find
for the off-diagonal component:  

      \begin{align}
      \label{eq:sigma_tr}
       \frac{1}{2\eta}\sigma_{\theta r}=-\frac{(l+2)}{r^{l+3}}a_lP_l^{1}
        +\frac{(l^2-1)}{lr^{l+1}}b_lP_l^{1}
             \end{align}  
     Concerning diagonal components, we only need the deviatoric stress $\Delta\sigma=\sigma_{r r}- \sigma_{\theta \theta}$, which is independent of the pressure. To first order it needs to be evaluated at $r=1$ and we get 
     \begin{align}
     \label{eq:delsigma}
       \frac{1}{\eta}(\sigma_{r r}- \sigma_{\theta \theta})=
       (l+1)\Big(3(l+2)a_l-3lb_l\Big)P_l\nonumber\\
       -\Big(a_l-\frac{(l-2)}{l}b_l\Big)P_l^{2}
             \end{align} 

             Eqs.\ref{stress} apply equally well to the interior flow field and yield 
 \begin{align}
 \label{eq:Sigma_tr}
       &\frac{1}{2\lambda\eta}\Sigma_{\theta r}  =  (l-1)r^{l-2}c_lP_l^{1}
        +\frac{l(l+2)}{l+1}r^{l}d_lP_l^{1},\\
        \label{eq:delSigma}
   &\frac{1}{\lambda\eta}(\Sigma_{r r}- \Sigma_{\theta \theta}) =
   \Big(3(l-1)c_l+3(l+1)d_l\Big)lP_l\\\nonumber
       &-\Big(c_l+\frac{(l+3)}{(l+1)}d_l\Big)P_l^{2}
 \end{align} 
for the corresponding stress components.

\section{Integrals}\label{appendix_integrals}
In this appendix, we summarise the integrals over products of
associated Legendre polynomials which are needed in the main text,
when projecting onto $P_1$:
\begin{align}
\label{eq:integral1}
&  \int_{-1}^{1}dx P_l^0(x)P_m^0(x)P_1^0(x)\nonumber\\
&  =\frac{2}{(2m+1)(2l+1)}\Big(
  (m+1)\delta_{l,m+1}+m\delta_{l,m-1}\Big),\\
\label{eq:integral2}
&  \int_{-1}^{1}dx P_l^1(x)P_m^1(x)P_1^0(x)\nonumber\\
&  =\frac{2l(l+1)}{(2m+1)(2l+1)}\Big(
  m\delta_{l,m+1}+(m+1)\delta_{l,m-1}\Big),\\
\label{eq:integral3}
&  \int_{-1}^{1}dx P_l^1(x)P_m^0(x)P_1^1(x)\nonumber\\
&  =\frac{2l(l+1)}{(2m+1)(2l+1)}\Big(
  \delta_{l,m+1}-\delta_{l,m-1}\Big),\\
\label{eq:integral4}
&  \int_{-1}^{1}dx P_l^2(x)P_m^1(x)P_1^1(x)\nonumber\\
&  =\frac{2(l-1)l(l+1)(l+2)}{(2m+1)(2l+1)}\Big(
  \delta_{l,m+1}-\delta_{l,m-1}\Big).
 \end{align}

\section{Droplet 2nd order}\label{app:droplet-second-order}
 We start from the four Eqs. (\ref{eq:secondOrderBCDrop}), multiply the first two by $P_1$ and the remaining 2 by $P_1^1$, and integrate over $z=\cos \theta$ to project onto the $l=1$ mode. The resulting $4\times 4$ system of linear equations takes on the form
 \begin{align}
 \label{eq:firstEqforU}
     -4a-2U & =3I_r^+,\\
  \label{eq:secondEqforU}    
     2c+2d-2U & = 3 I_r^-,\\
 \label{eq:thirdEqforU}
     4a-4c-8d & = 3 I_\theta,\\
\label{eq:fourthEqForU}
     -8\eta^+ a- 4\eta^-d & = I_\sigma,
 \end{align}
with
 \begin{align}
 \label{eq:Ir}
     I_r^\pm & = \int_{-1}^1 dz \left(v_\theta^\pm f' - f\partial_rv_r^\pm\right)P_1,\\
     \label{eq:Itheta}
     I_\theta & = \int_{-1}^1 dz  f\big[\partial_rv_\theta\big]P^1_1,\\
     \label{eq:Isigma}
     I_\sigma &= \int_{-1}^1 dz \left(f'\big[\Delta\sigma\big] + f\big[\partial_r\sigma_{r\theta}\big]\right)P_1^1.
 \end{align}
 Here, $\bm{v}^+=\bm{v}$ and $\bm{v}^-=\bm{V}$. 
The term $\Delta\sigma$ represents the deviatoric stress, defined as $\sigma_{rr}-\sigma_{\theta\theta}$, while the angular brackets signify the discontinuity across the surface of the unit sphere, that is, $\big[A(r, \theta)\big]=A(1-\epsilon, \theta)-A(1 + \epsilon, \theta)$ for $\epsilon \to 0_+$. The velocity and the stress fields in the inhomogeneities are obtained from the first-order solutions given by Eqs. (\ref{eq:firstorderad}, \ref{eq:firstorderc}, \ref{eq:firstordera}) and Eqs. (\ref{eq:sigma_tr} - \ref{eq:delSigma})).  The swimming velocity $U$ is obtained from the system \ref{eq:firstEqforU}-\ref{eq:fourthEqForU} in the form
 \begin{align}
     U=-\frac{3\eta^+I_r^+ + 3\eta^-I_r^- + \frac{3}{2}\eta^-(I_r^+ + I_\theta) - \frac{1}{2} I_\sigma}{3\eta^-+2\eta^+}.
 \end{align}

To evaluate the four inhomogeneities $I^\pm_r, I_\theta, I_\sigma$ we insert the expansions Eqs. (\ref{eq:vrexp}, \ref{eq:vtexp}, \ref{eq:Vrexp}, \ref{eq:Vtexp}) and Eqs. (\ref{eq:sigma_tr}-\ref{eq:delSigma})  in Eqs. \ref{eq:Ir}-\ref{eq:Isigma} and perform the integrations with the help of Eqs. (\ref{eq:integral1}-\ref{eq:integral4}).  As an example, consider $I_r^\pm$. It consists of a double sum of terms arising from $v_r=\sum_l v_{r,l}P_l$, from $v_\theta=\sum_l v_{\theta,l}P^1_l$ and from $f=\sum_l f_l P_l$. It takes on the form
\begin{align}
    I^\pm_r=\sum_{l, l'} \int_{-1}^{1} \Big(f_{l'}v^\pm_{\theta, l} P_l^1P_{l'}^1P_1-f_{l'}\partial_rv^\pm_{r,l} P_lP_{l'}P_1\Big).
\end{align}
 We insert Eqs. (\ref{eq:integral1}) and (\ref{eq:integral2}) and perform the summation over $l'$. The remaining $l$-series can be written in the form
\begin{align}
    I^\pm_r &= \sum_l B_l\left(f_l v^\pm_{\theta, l+1} + f_{l+1}v^\pm_{\theta, l}\right) \\\nonumber
    & - \sum_l A_l\left(f_l\partial_r v^\pm_{r, l+1} + f_{l+1}\partial_rv^\pm_{r,l}\right) 
\end{align}
with
\begin{align}
    A_l & = \frac{2(l+1)}{(2l+1)(2l+3)},\\
    B_l & = l(l+2)A_l.
\end{align}
The other terms are treated in the same way. $I_\theta$ contains terms $\sim \int P^1_lP^1_{l'}P_1$ and gives the following result:
\begin{align}
    I_\theta &= \sum_l A_l\left(f_l\big[\partial_rv_{\theta, l+1}\big](l+2)- f_{l+1}\big[\partial_rv_{\theta, l}\big] l\right).
\end{align}

The inhomogeneity $I_\sigma$ contains terms $\sim \int P^1_lP_{l'}P^1_1$ from $f\partial_r\sigma_{\theta r}$. The contributions from $f'\Delta\sigma$ produce terms $\sim \int P_lP^1_{l'}P^1_1$ (denoted by $f'\Delta\sigma^{(0)}$) and terms $\sim \int P^2_lP^1_{l'}P^1_1$ (denoted by $f'\Delta\sigma^{(2)}$), which can be read from Eq.(\ref{eq:delsigma} and \ref{eq:delSigma}). The resulting l-series takes on the form    
\begin{align}
    I_\sigma & = \sum_l A_l\left(f_l \big[\partial_r\sigma_{\theta r, {l+1}}\big](l+2) - f_{l+1}  \big[\partial_r\sigma_{\theta r, {l}}\big]l\right)\\\nonumber
     & + \sum_l A_l\left(f_{l+1}\big[\Delta\sigma^{(0)}_l\big](l+2) - f_{l}\big[\Delta\sigma^{(0)}_{l+1}\big]l\right)\\\nonumber
     & + \sum_l B_l\left(f_{l}\big[\Delta\sigma^{(2)}_{l+1}\big](l+3) - f_{l+1}\big[\Delta\sigma^{(2)}_{l}\big](l-1)\right).   
\end{align}
All first-order quantities are proportional to $\dot{f}_l$, so that the result for $U$ can be written in the form of Eq. (\ref{eq:Uparticle}) in the main text. 

The explicit calculation of the terms $R^{drop}_l, S^{drop}_l$   and $\overline{U}_l$ is straightforward but tedious, if not done with the help of computer algebra.  
The polynomials $r_{l}^{(n)}(\lambda), s_{l}^{(n)}(\lambda)$ in Eqs. (\ref{eq:Rdrop}, \ref{eq:Sdrop},) take on the form
\begin{align}
    r_{l}^{(4)} & = -4(\lambda + 1)^2,\\\nonumber
    r_{l}^{(3)} & = -6(\lambda +1) (3\lambda +5),\\\nonumber
    r_{l}^{(2)} & = -2(31\lambda^2 + 83\lambda +46),\\\nonumber
    r_{l}^{(1)} & = -3(29\lambda^2 + 69 \lambda +28),\\\nonumber
    r_{l}^{(0)} & = -9(\lambda+1)(3\lambda +2),
\end{align}
and 
\begin{align}
    s_{l}^{(4)} & = 4(\lambda + 1)^2,\\\nonumber
    s_{l}^{(3)} & = 2(\lambda +1) (\lambda +7),\\\nonumber
    s_{l}^{(2)} & = -2(26\lambda^2 + 25\lambda +5),\\\nonumber
    s_{l}^{(1)} & = -(65\lambda^2 + 169 \lambda +44),\\\nonumber
    s_{l}^{(0)} & = -12\lambda(2\lambda +7).
\end{align}
These expressions complete the analytical solution of the drop velocity.

\section{Energy dissipation}\label{app:dissipation}
For the outer flow fields Eqs.(\ref{eq:vrexp}, \ref{eq:vtexp}) and the inner flow fields Eqs. (\ref{eq:Vrexp}, \ref{eq:Vtexp}), the integral Eq.(\ref{eq:dissipation_integral}) can be evaluated by inserting the first-order solutions.
For outer flow one gets 
\begin{align}
\label{eq:dissipation_out_rr}
 & \int_{\partial V}d^2x \; \sigma_{rr}^{dis}v_r= \sum_l D_{r,l}^{out} \dot{f}_l^2=\nonumber\\
  & 8\pi\eta\sum_{l\geq 2}\frac{(l+1)^2}{2l+1}(-a_l+b_l)((l+2)a_l-lb_l),\\
\label{eq:dissipation_out_rth}  
 & \int_{\partial V}d^2x \; \sigma_{r\theta}^{dis}v_\theta=\sum_l D_{t,l}^{out} \dot{f}_l^2\nonumber\\
                        &8\pi\eta\sum_{l\geq 2}\frac{(l+1)l}{2l+1}(a_l-\frac{l-2}{l}b_l)\big(-(l+2)a_l+\frac{l^2-1}{l}b_l\big).  
\end{align}

The energy dissipation for the inner space can equally well be computed and results in
\begin{align}
\label{eq:dissipation_in_rr}
 & \int_{\partial V}d^2x \; \Sigma_{rr}^{dis}V_r = \sum_l D_{r,l}^{in} \dot{f}_l^2\nonumber\\
  &= 8\pi\eta\sum_{l\geq 2}\frac{l^2}{2l+1}(c_l+d_l)((l-1)c_l+(l+1)d_l),\\
  \label{eq:dissipation_in_rth}
 & \int_{\partial V}d^2x \; \Sigma_{r\theta}^{dis}V_\theta  \sum_l D_{t,l}^{in} \dot{f}_l^2\nonumber\\
 &=8\pi\eta\sum_{l\geq 2}\frac{(l+1)l}{2l+1}(c_l+
   \frac{l+3}{l+1}\hat{d}_l)((l-1)c_l+\frac{l(l+2)}{l+1}d_l). 
\end{align}
The flow inside the vesicle can be calculated in analogy to the ouside flow and yields
\begin{equation}
  d_l=-\frac{l-1}{2l}\dot{f}_l, \quad c_l=\frac{l+1}{2l}\dot{f}_l.
\end{equation}
After inserting the first-order solutions into these expressions, one gets the Eqs. (\ref{eq:dissipate_particle}-\ref{eq:dissipate_drop}). For the droplet, the dissipation coefficients take on the explicit form
\begin{align}
    D_{r,l}^{out} = -8\pi\eta \frac{l(\lambda +1) - \lambda +2}{(\lambda +1)(2l+1)^2},\nonumber\\
    D_{t,l}^{out} = -36\pi\eta\frac{(l+1)\lambda^2 + l\lambda}{(\lambda+1)^2(l+1)^2(2l+1)^2},
\end{align}
and
\begin{align}
    D_{r,l}^{in} = \lambda D_{r,l}^{out},\nonumber\\
    D_{t,l}^{in} = D_{t,l}^{out}.
\end{align}

\section{\label{app:swim-strokes} Swim strokes with fixed surface area}
With the designation
\begin{equation}
    g_l := \frac{F_l^2}{\Delta}\frac{(l+2)(l-1)}{(2l+1)}
\end{equation}
we find constraint equations from Eq(\ref{eq:area-constraint}) and Eq.(\ref{eq:simpleHarmonicDeform}), which take on the simple form
\begin{align}
\label{eq:constraint1}
    & \sum_{l\geq 2} g_l  = 1,\\
    & \label{eq:constraint2}
    \sum_{l\geq 2} g_l e^{2i\alpha_l}  = 0.
\end{align} 
We restrict the deformations to the three $l$ components $l=2,3$ and $4$.  For $F_4\neq 0$, we eliminate $g_4=1-g_2-g_3$ in Eq. (\ref{eq:constraint2}) and write the remaining equations in the form
\begin{align}
\label{eq:constraint3}
    2\hat{g}_3\sin \alpha_3 & = 1-2 g_2\sin^2\alpha_2,\\
    \label{eq:constraint4}
     2\hat{g}_3\cos\alpha_3 & = -2g_2\sin\alpha_2\cos\alpha_2,
\end{align}
with $\hat{g}_3=g_3\sin\alpha_3$.  First, we square and then add the two equations (\ref{eq:constraint3}) and (\ref{eq:constraint4}), and so we can express $\hat{g}_3$ in terms of $g_2$ and $\alpha_2$. Then we can obtain $\sin{\alpha_3}$ and $\cos{\alpha_3}$ from Eqs. (\ref{eq:constraint3}) and (\ref{eq:constraint4}) and finally, $g_3=\hat{g}_3/\sin\alpha_3$. Note that the solutions of Eqs. (\ref{eq:constraint3}) and (\ref{eq:constraint4}) remain unchanged for $\alpha_2 \to \alpha_2 +\pi$.  They can be written in the form  
\begin{align}
    g_3 & = \frac{1}{2}\frac{C}{1-2g_2\sin^2\alpha_2},\\
    \cos\alpha_3 & = \frac{-2g_2}{\pm C^{1/2}}(\sin\alpha_2\cos\alpha_2),\\
    \sin\alpha_3 & = \frac{1}{\pm C^{1/2}} (1-2g_2\sin^2\alpha_2),
\end{align}
with $C=1+4(g_2^2-g_2)\sin^2\alpha_2$.  The phase $\alpha_3$ is determined only up to an angle $\pi$, due to the two roots $\pm C^{1/2}$. In the following, we discuss only the $+$-branch. Changing $\alpha_3\to \alpha_3 + \pi$ changes the sign of the swimming velocity $U$. Thus, it should always be remembered that for every motion discussed in the following, there is a corresponding motion in the opposite direction. The values of $g_2$ must be restricted to $0 < g_2 <1/2$ (corresponding to $0 < F_2< F_{2, max}=\sqrt{5\Delta/8}$) to ensure that $g_3$ and $g_4$ are positive and $|\cos\alpha_3| < 1$. 

The special case $F_4=0$ has only 3 parameters $F_2, F_3, \alpha_2$ In this case $g_2, g_3$ and $\alpha_2$ are completely determined to $g_2=g_3=1/2, \alpha_2=\pi/2, 3\pi/2$, and no adjustable parameters remain.

\section*{Author contributions}
Both authors contributed equally.

\section*{Conflicts of interest}
There are no conflicts to declare

\section*{Data availability}
All data is contained in the article.

%\section*{Acknowledgements}
%The Acknowledgements come at the end of an article after Conflicts of interest and before the Notes and references.

%%%END OF MAIN TEXT%%%


\balance

\renewcommand\refname{References}

%%%REFERENCES%%%
\bibliography{Amoeboid.bib} %You need to replace "rsc" on this line with the name of your .bib file
\bibliographystyle{rsc} %the RSC's .bst file
\end{document}